# Neural Replicator Analysis
# for virus genomes binomial systematics in metagenomics

Alexandr A. Ezhov

Troitsk Institute for Innovation and Fusion Research, 108840, Troitsk, Moscow, Russia
ezhov@triniti.ru

**Abstract**

We have presented some arguments to substantiate the usefulness of neural replicator analysis (NRA) for constructing variants of the natural binomial classification of virus genomes based only on knowledge of their complete genomic sequences, without involving other data on the phenotype, functions, encoded proteins, etc., and also without the need of genomic sequences alignment. Perhaps this will make sense when processing metagenomic data. This makes it possible to construct the binomial classification accepted for the viruses themselves. We restrict ourselves to three families of viruses having dsDNA circular genomes (*Papillomaviridae*, *Polyomaviridae* and *Caulimoviridae*) and partly to the family *Geminiviridae* having ssDNA genomes though the approach presented can be also applied to genomes of other dsDNA, ssDNA and ssRNA viruses, including linear ones (some results for *Mitoviridae* are also presented). It is argued that binomial classification of virus genomes which is difficult to apply in all cases can nevertheless be informative tool of revealing virus properties, areal of hosts, forms of diseases and can also show the connections of the viruses belonging to different families and even to different kingdoms.

*Keywords:* metagenomics, neural replicator analysis, binomial virus genome classification

**Introduction**

Today, most viruses are known solely from sequence data obtained from metagenomic studies, and these metagenomics sequence data are used by the International Committee on Taxonomy of Viruses (ICTV) to develop an official taxonomy scheme for viruses without the use of additional phenotypic characteristics. This differs from the earlier approach, which used information about host range, replication cycle, structure and properties of viral particles, etc. to determine viral groups [1]. Despite of the presence or absence of phenotypic data, taxonomic categories are organized into *a hierarchy* that includes species, genera, families, and orders of viruses [2]. According to [3, 4], viruses are classified in 9 orders, 131 families, 46 subfamilies, 803 genera and 4 853 species, based on the results of the ratification vote after the 49th ICTV Executive Committee meeting. But, as Simmond and Aiewsakun note that [3]:

> "… unified virus taxonomy is a rather ramshackle construction, with taxonomic assignment rules often being based on quite different and inconsistent criteria between virus groups".

Indeed, the classification of viruses faces many problems, such as, for example, the definition of *species*. In 1991 ICTV changed the definition of virus species, accepting the formulation given by Marc Van Regenmortel:

> *"A virus species is a polythetic class of viruses that constitute a replicating lineage and occupy a particular ecological niche"* [2, 5].

A polythetic class consists of members that have a number of common properties, but not all have one common property [2]. This definition is flexible: as Simmond and Aiewsakun wrote in [3]:

> *"Flexibility in what defines species is implicit in the polythetic species definition developed by Marc Van Regenmortel, in which constellations of properties, none of which would be individually essential for species inclusion or exclusion, are formulated to produce a highly intuitive and effective descriptive definition."*

Note that this definition is in full accordance with the attractor neural network model of content-addressable memory [6], which, can be used to place different patterns in one class that do not have a common property, but have a common *prototype* [7]. Such networks are suitable for virus classification, especially in the situation of variability of their genomes. But the above definition of virus species corresponding to the lowest taxonomic level in the hierarchy was changed and approved by ICTV in 2013 and now has the form [8, 9]:

> *"A species is a monophyletic group of viruses whose properties can be distinguished from those of other species by multiple criteria."*

This revised definition of a viral species includes a reference to evolutionary history − it defines a viral species in terms of monophyly: a monophyletic taxon is defined as a group composed of a set of organisms, including the most recent common ancestor of all those organisms and all the descendants of the most recent common ancestor.

This definition has been widely criticized for not mentioning the concept of a polythetic class, for example in [10]:

> *"…by defining a species as a monophyletic group of viruses, the authors were unable to provide a practical criterion for distinguishing one monophyletic species from another since every species, genus and family can be considered to be a monophyletic class".*

> *"Another reason monophyly is not a generally applicable criterion for species demarcation is the common occurrence in many viruses of recombination and reassortment phenomena among parts of virus genomes and of exchanges of genes between viruses and their hosts. This produces chimeric viruses with polyphyletic genomes* [11] *and it is then logically impossible to accurately represent such multi-dimensional phylogeny in a monophyletic scheme* [12]*".*

and in [13]:

> *"The existence of prolific horizontal genetic transfer among various groups of viruses presents a challenge to this definition"*

But, despite this controversy, the problem of virus species definition remains unsolved in the metagenomic era. As Simmond and A Aiewsakun stated [3]:

> *"... you can't use descriptive species definitions for viruses where there is nothing to describe except its nucleotide sequence".*

Nevertheless, species demarcation is essentially based on the comparison of vials genomes, which requires alignment of the nucleotide or corresponding amino acid sequences for the complete genome or its parts – genes encoding important proteins. For example, this can be done for RNA-dependent RNA polymerase (RdRP). In this case, it is claimed that new species differs by more than 10% difference in amino acid sequence with known ones. Sequence comparison has traditionally been performed by various alignment-based methods. These methods often attempt to maximize the alignment score calculated as the sum of substitution scores minus gap penalties. Popular algorithms include Smith-Waterman, Needleman-Wunch, Muscle, and others [14,15]. But there was no fixed sequence divergence threshold that would identify members of the same species. Sometimes individual species are distinguished on the basis of their differences in geographic range, host associations, and pathogenicity rather than their genetic relationship. There are also methods without alignment based on comparison of sequence correlation coefficients [16].

What is essential – all these methods are used to construct taxonomic hierarchical trees, reflecting an evolutionary approach to virus taxonomy. But a biological classification can be built both without reference to evolution, and it does not have to be hierarchical. The use of a tabular form to represent biological objects instead of hierarchical trees was discussed in the studies of Alexander A. Lubischew [17]. Lubischew argued that a natural system in biology can have a form that does not reflect the evolution of species, but can have a form similar to Mendeleev's periodic table of chemical elements. A similar form of "*periodic table of cell types*" was recently proposed by Bo Xia and Itai Yanai [18].

The goal of this paper is to prove that Neural Replicator Analysis (NRA) introduced in [19] makes it possible to construct a tabular binomial form for virus genomes that have some properties of natural system of genome sequences. We hope that this approach can also be used to classify [20]. Note, that the change approved in 2021 by ICTV is the adoption of a uniform binomial format for naming of virus species [21]. The form of classification table form also makes it possible to characterize the genomes of viruses using two coordinates determine the position of the genome in the table.

The structure of the paper is as follows. First, we present basic concepts of NRA introduced in [19] and applied to the study of viroid circular RNA. Then, in section 2 we apply this approach to the analysis of human papillomaviruses and demonstrate, first of all, its ability to separate viruses that damage mucosal and cutaneous tissues. Based on the results of this analysis, the bottom row of the table corresponding to the absence of replicators for KM-encoded genomes [19] begins to form. Section 3 presents the NRA of polyomaviruses. The two new cells in the upper rows are begin to fill with genome data, in particular, avian viruses. In Section 4, members of *Caulimoviridae* family are studied and it is shown, that the genomes of the genus *Badnavirus* fills the cells of a certain column of the table. After that, in Section 5, some members of the *Geminiviridae* and *Mitoviridae* families are studied and the first joint table of the virus genome is presented. The last section presents some conclusions and a discussion of the results.

## 1. Neural Replicator Analysis

The basic artificial neural network model used in NRA is the self-reproducible neural network (neural replicator) [19,22,23]. This model includes the mechanism of synchronously changing threshold of all neurons having binary states $x_i$ (+1 or −1) in the standard Hopfield network [6]. It is suggested that ancestor Hopfield network has arbitrary matrix of interconnections and corresponding set of attractors (stable states) for zero neuron thresholds. This network is placed in a *network ensemble* (e.g., one or two dimensional) consisting of the untrained networks having zero synaptic matrix. The ancestor network can force neighbor network neurons to take values of their neuron states in through one-to-one interconnections in the course of information transmission [22]. The signal of the *start of this transmission* arises when ancestor network puts all the thresholds of their neurons to the very low negative values at once. In this case all states of ancestor network neurons take maximal values (+1). This maximally excited state of ancestor network *opens the channel* of information transmission to neighbor network. Then all thresholds of ancestor network start to grow synchronously taking the same values. At some threshold level the state of some neurons *become unstable* and neural dynamics starts until *equilibrium state* at this threshold will be reached (note, that threshold grow is very slow to permit this process to terminate). This equilibrium (stable) state is transmitted to the neighbor network *forcing it to learn this pattern* with Hebbian rule [6]. Then the growth of thresholds in ancestor network continues and it transmits its quasi stable attractors arisen at different threshold levels to a neighbor network which learns all of them. When the threshold level becomes high enough all neurons become passive (their states take values $x_i = -1$) and this passive network state is interpreted by neighbor network as the signal of the finish of information transmission. After this course the neighbor network learns all quasi stable (stable in given threshold interval) states of ancestor network and becomes a new ancestor network able to transmit information to its untrained neighbor network. So, for example, in linear chain of networks a one-directional wave of learning can be organized. The remarkable phenomena observed in such a system [19,22,23] is that after few steps of transmission a special network arises in a chain which transmit further *just those patterns which it learned from its neighbor*. In other words, this network produces its *exact copy*, or is self-reproducible. In effect, identical networks arise and spread through the system. The self-reproducible networks are absolutely transparent ones − they show as quasi attractors all learned patterns during the cycle of threshold growth.

*Incomplete codes of nucleotide sequence*

The model suggests that neurons take binary values. Though many generalizations of this model permit to avoid this restriction just such code scheme was used for genomic analysis in a previous paper [19]. In this paper *non-traditional representation* of nucleotide sequences was used. Instead of four-letter genetic code *two binary code schemes* to represent these sequences were introduced. The first code (called WS code) combines the Watson-Crick pairs (AT) and (CG) and presents them as a weak (AT) pair encoded by "–1" and a strong (CG) pair encoded by "+1". The second keto-amino (KM) code combines a wobble pair (TG) encoded with "+1" and a less stable (AC) pair encoded with "–1". For example, the sequence (ATACGGGCTGAA) will be represented by two binary strings: (−1−1−1 +1+1+1+1+1 −1+1 −1−1) (WS code) and (−1+1−1−1+1+1+1−1+1+1−1−1) (KM code).

*Replicator Tables*

These two incomplete codes were used to construct sets of networks of different sizes *K* (starting from 3) with the Hebbian interconnections calculated with the use of patterns generated by sliding the nucleotide sequence consisting of *N* nucleotides with a window having a length *K* (Fig.1 – [19]). *N* resulting patterns (note, that their number does not depend on *K*) are used to form the Hebbian matrices of interconnections of the two parent fully connected Hopfield networks (for WS- and KM-encoded patterns, correspondingly). Then self-reproducible replicators were obtained according to the procedure described above. For simplicity and to avoid the ambiguity (the appearance of different sets of replicators) asynchronous but ordered dynamics for updates of the states of neurons in the Hopfield network is suggested. The results of studies are presented in Replicator Table (RT )[19] which presents the presence or absence replicators for both code schemes (WS and KM) and different network size, *K* (Fig.1, right).

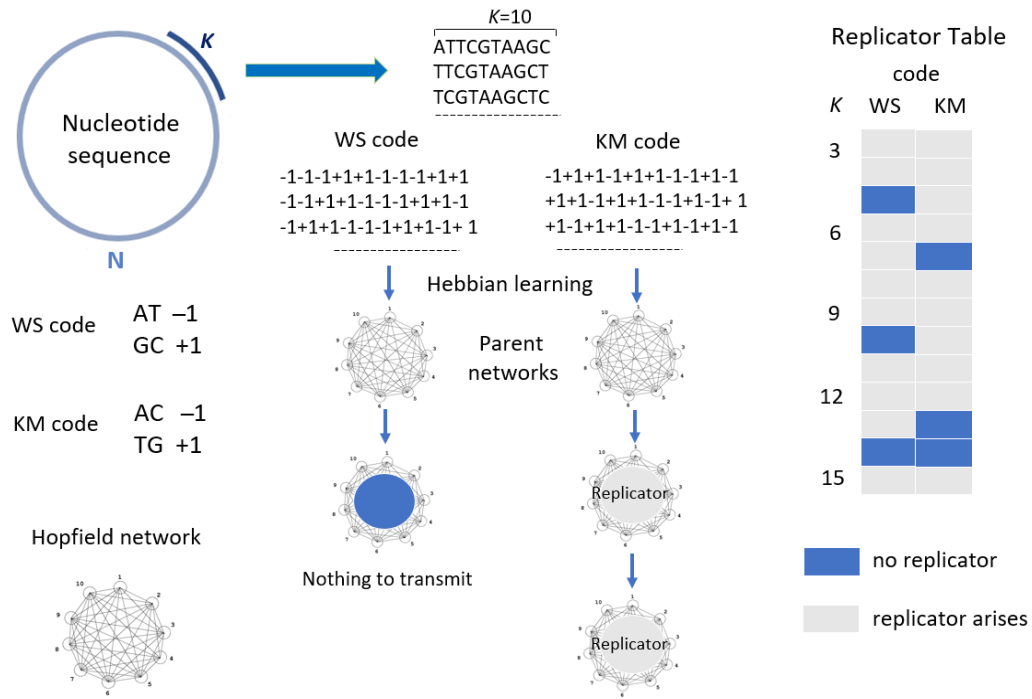

**Fig. 1.** The scheme for analysis of nucleotide sequence. A circular nucleotide sequence containing *N* nucleotides is traced by sliding a window having a length *K* with increment of one nucleotide. Two codes (WS and KM) are used to convert each *K*-nucleotide fragment into two binary strings. Then *N* resulting patterns (their number does not depend on *K*) are used to form the Hebbian matrices of interconnections of the two parent fully connected Hopfield networks. Then the iterative procedure for increasing the threshold of neurons and rewriting the detected stable states into the descendant networks in a linear network chain is continued until a final replicator network arises or an empty set of patterns for transmission is formed in the final network. Here and further the first case is represented by a gray box in the Replicator Table (RT), while the case without a replicator is represented by a blue box. Note, that the absence of replicator corresponds to the generation of network which starting from the state of all active neurons at the lowest threshold value in reaching a definite higher threshold value goes directly to the state where *all* neurons become passive. The first and the second (maximally and minimally excited) states define non-specific "epileptic" and "dead" patterns of network activity and are not included into the set, which can be further transmitted, but only serve as signals of the beginning and the end of information transmission. In the latter case no patterns for transmission are generated and replicator does not arise. This procedure is performed for different lengths of sliding window *K*: starting from *K*=3 [19].

The remarkable finding was as follows: sets of replicators *differ substantially* for two incomplete representations of the viroid RNA sequences (obtained using WS and KM codes). The simplest difference is the fact that for a sliding window of the same size the source parent network can generate a nontrivial replicator with a non-empty set of the patterns for transmission, or non-replicating network with empty set of patterns for transmission. This last network cannot generate descendants or, in other words, cannot breed. Further we will see that the last situation is rather common for some virus types, but in general virus RTs have non-trivial form.

In [19] it was demonstrated that despite a wide range of RT forms some reasonable approximate categorization of two viroid families can be derived. For example, *Avsunviroidae* family of viroids can be characterized by the absence of replicators having a short length (up to 6-8) for the KM-code, etc.

*Fuzzy motifs*

Other interesting phenomenon is connected to the replicator transmitted patterns - fuzzy motifs. It was shown that patterns transmitted by replicators contain additional information and often have interesting symmetries and periodicities [19]. More details about the different additional results of the application of NRA to the study of rather short viroid genomes are also presented in [19].

*Application of NRA to virus genomes analysis*

Obviously, the approach proposed in [19] and applied to the analysis of viroids can also be applied to virus genomes. Hepatitis delta virus (HDV) has the smallest DNA genome, closely resembling the RNA genome of viroids [24], and its replicator table has a form typical of viroids as also of some narnaviruses and mitoviruses (Fig. 2).

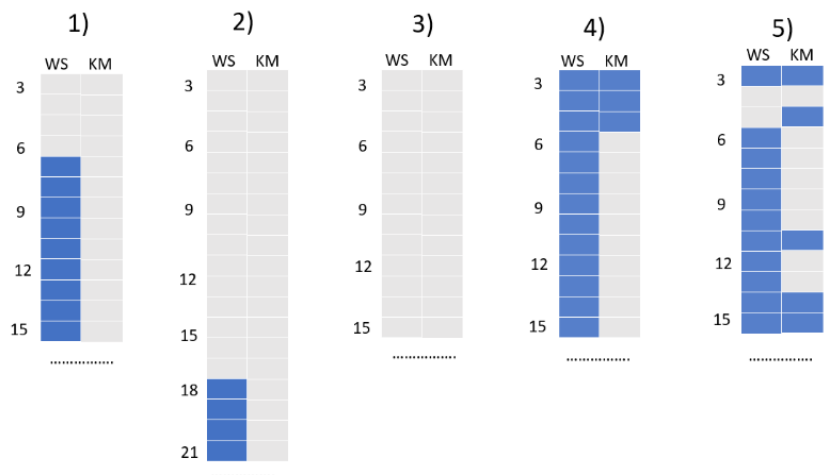

**Fig. 2.** The examples of virus genomes RTs, where both WS- and also KM- encoded sequences generate self-reproducible neural networks – replicators (gray boxes). The absence of replicators is illustrated by blue boxes. 1) Hepatitis delta virus genomic RNA, GenBank: D01075.1M, 1682 bp, 2) Saccharomyces 23S RNA narnavirus NC_004050 2891 bp, 3) Fusarium poae narnavirus 1 NCBI Reference Sequence: NC_030865, 2297 bp, 4) Ophiostoma mitovirus 5 NC_004053.1 2474 bp 5) Binucleate Rhizoctonia mitovirus K1 isolate NC_027921.1 2794 bp

RTs of other hepatitis viruses have different shapes, but, as we will further see, the RTs of hepatitis A, C, and E viruses have forms of RT typical for human papillomaviruses. The main feature of these viruses is that they do not have replicators generated on KM-encoded DNA or RNA sequences (Fig. 3).

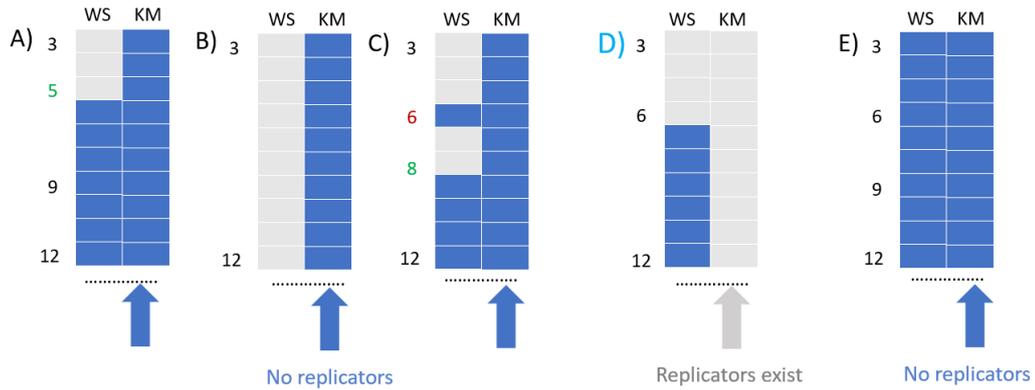

**Fig. 3.** RTs of hepatitis viruses. With the exception of delta viruses (D), all other viruses (A,B,C,E) are characterized by the absence of replicators built using KM-encoded genomic sequences (only blue boxes in the right columns). A) Hepatitis A isolate p16 virus genomic RNA, GenBank: KP879217.1, 7476 bp, B) Hepatitis B virus isolate MT, GenBank: KC492739.1, 3215 bp, C) Hepatitis C virus genotype 1, NCBI Reference Sequence: NC_004102, 9646 bp, D) Hepatitis delta virus genomic RNA, GenBank: D01075.1M, 1682 bp, E) Hepatitis E virus, NC_001434, 7176 bp.

Consider RT of the hepatitis E virus (Fig. 3 E). It also lacks replicators when its RNA sequence is represented by the WS code. For hepatitis A and C, these replicators exist, but only up to a certain maximum size of the neural network: five for hepatitis A and eight for hepatitis C. Also for hepatitis A, there are all replicators of smaller network dimensions (starting from 3). We will call such RTs monotonic. In contrast, for the hepatitis C virus, there is no replicator for network size 6. We will call corresponding RTs non monotonic and will further use asterisks to mark corresponding virus data.

For the case of human papillomaviruses considered in Section 2 we can forget about right columns of RT (replicators for KM-encoded genomes do not exist) and use only maximal size of replicators corresponding to WS code for the analysis. Note, that the same situation arise for the virus SARS 2 isolate 2019-nCoV (WHU01 29881 bp ssRNA(+)) which RT is the same as the RT for hepatitis A. But in similar cases we can use additional information related to the patterns which use replicator networks for information transmission. For example, for hepatitis A replicators of the size 5 are built on only one pattern: (1 1 -1 1 1), while for SARS virus on two patterns: (1 1 -1 1 1) and (-1 1 -1 -1 1). As we will see the forms of RT and forms of patterns can give us interesting information about virus genomes similarities and also about their divergence.

## 3. Neural Replicator Analysis of human papillomaviruses

Here, we apply the approach described above and used in [19] to the analysis of human papillomavirus (HPV) genomic sequences. HPVs are small, non-enveloped, double-stranded DNA viruses belonging to the *Papillomaviridae* family [25]. The taxonomy of these viruses is usually based on the study of the nucleotide sequences of the main viral capsid protein L1 [26]. HPV types belonging to different genera share less than 60% similarity within the L1 portion of the genome. Different types of viruses within the genus have 60 to 70% similarity. The new HPV type has less than 90% similarity to any other HPV type. The papillomavirus nomenclature at the species level and above is determined by the papillomavirus study group of ICTV [27]. Human papillomaviruses are classified into 5 genera − $\alpha, \beta, \gamma, \mu$, and $\nu$, containing many species and types: the number of these types increases linearly with time for genus $\beta$ and extremely rapidly for the genus $\gamma$ − the rate of detection of HPV types increases, mainly as the result of metagenomic sequencing [28]. Here we use species and types taxonomy data provided in [29] and the relevant NCBI and GenBank references are provided in the Materials section.

Instead of RTs, which in this case do not have replicators for KM-encoded sequences for a genome size of about 8000, we will use a convenient visualization of the situation, showing only replicators with WS-code. Next, we will use colors to mark the maximum size, $N_{max}$, of the replicator neural network generated using the WS-encoded genome sequence. Thus, the situation of the absence of replicators will be marked in black, the presence of only a replicator of size 3 in purple, the presence of a maximum size 4 in blue, the maximum size 5 in light blue, the maximum size 6 in green, the maximum size 7 in yellow, the maximum size 8 in orange and the maximum size 9 in red (Fig. 4). It turns out that this set of colors is sufficient to characterize all replicators of maximum size for all types of human papillomaviruses studied. We will also use one or two asterisks to denote cases with non-monotonic sets of replicators when a replicator does not exist for one or two smaller than maximal size, respectively. We start with the genus *Alphapapillomavirus* and present the results of their study in Fig. 4.

Genus*: Alphapapillomavirus*

Standard characteristics of this genus is that "*Alpha HPVs preferentially infect the anogenital and oral mucosa, causing both malignant and benign neoplasms. Cutaneous lesions have also been observed*" [30]. One of oncological disease connected with α-genus is the cervical cancer (note, however, that HPV belonging to $\beta$ and $\gamma$- genera are also considered as carcinogenic cofactors of cervical cancer [31])

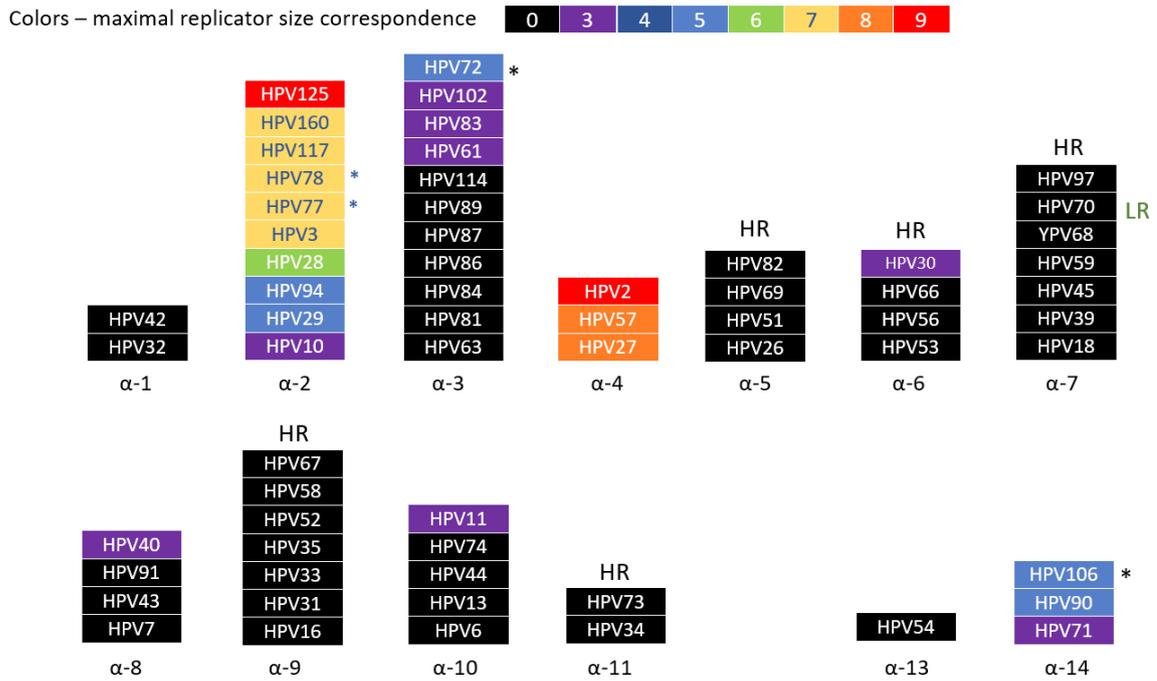

**Fig. 4.** The maximal size of replicators for different species and types of the *α* genus of human papillomaviruses. The sets of replicator patterns of the types HPV77, HPV78, HPV72 and HPV106 are non-monotonic (replicator of one of the size lower than the maximal size does not exist – corresponding colored boxes are marked by the asterisk)

There are some interesting observations that can be seen in this picture. First, the *α*-2 and *α*-4 species, which have a large size of replicators (up to 9), differ significantly from other species of the genus *α*. It is noteworthy that, unlike the types of other species, types of these two species cause the formation of skin warts (genus *α* also includes a few cutaneous HPV types (HPV2, 3, 7, 10, 27, 28, and 57), which cause common and plantar warts. Also, most types of high oncogenic risk (VR) are characterized by the absence of replicators (black boxes). Thus, using NRA, we can recognize a clear division of the genus *α* a into two subgenera, which was not obtained by a method based on the study of the similarity of the nucleotide sequences of the main capsid protein L1 [26].

More information can be obtained by considering patterns which are transmitted by replicators of maximal size. In all cases they transmit single patterns which for all types of species *α*-2, *α*-4, and also *α*-3 are *periodic* with period equals to 2 (Fig. 5). With only one very interesting exception (which will be further discussed) such periodic transmitted patterns are typical only to genus *α*. However, for species *α*-14 non-periodic patterns are transmitted by replicator of the size 5. In order to clarify the situation with *α* -14 let us consider genus *β* of human papillomavirus.

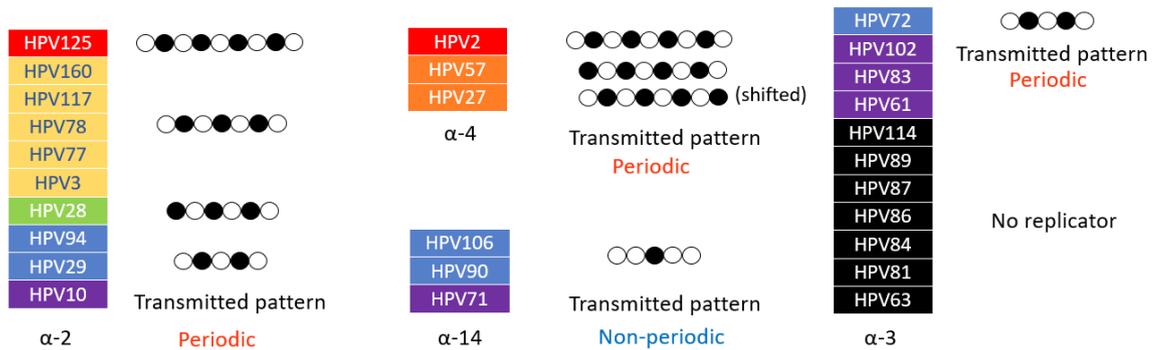

**Fig. 5.** The single patterns transmitted by replicator networks of maximal size are presented. Neuron states (pattern binary components) are represented by black (state equals to –1) or white (+1) circles. Transmitted patterns corresponding to the *α*-2,3,4 species are periodic with period equals to 2 (note, that patterns of HPV27 and HPV57 are complementary – shifted by one position). On contrary, the pattern transmitted by replicators corresponding to types HPV90 and HPV106 belonging to *α*-14 is not periodic.

Genus: *Betapapillomavirus*

*β*-HPVs cause only skin lesions and exist in a latent form in the general population, but are activated under conditions of immunosuppression [30]. *β*-HPV types under the influence of certain cofactors can also trigger a malignant process. Recent studies point to the role of human papillomavirus *β*-types and HPV-associated inflammation in the development of squamous cell skin cancer (the second most common non-melanoma skin cancer after basal cell carcinoma). But *β*-HPV infection appears to play an important role in initiating carcinogenesis, but not in tumor progression [32]. NRA shows that the "*coloring*" of *β*-papillomaviruses differs from what we observe for *α*-papillomaviruses (Fig. 3). It is characterized by a dominance of replicators with a maximum size of 5 (blue boxes), a lack of larger replicators (such as those of *α*-2 and *α*-4), and a small number of types without replicators at all (Fig. 6). Even more remarkable, all 5-neuron replicators transmit a single pattern that is *identical* to the non-periodic pattern of HPV90 and HPV106 types belonging to species *α*-14 (Fig. 4). So, in terms of NRA, should species *α*-14 be moved to genus *β* or other genera? We can clarify this by looking at the *γ*- and *μ*- genera (we realize that this analysis is rather rough and does not claim to draw any solid conclusions). We also note that *β*-1 is the only species of human papillomaviruses containing types HPV8, HPV47, HPV99, which have transmitted patterns of length 4, and type HPV8 has a complex set of such patterns, including not one, but four members. This type is unique among all human papillomaviruses as also HPV28 (*α*-2) which has pattern length of 6 and is associated with a high risk of developing squamous cell skin cancer [32].

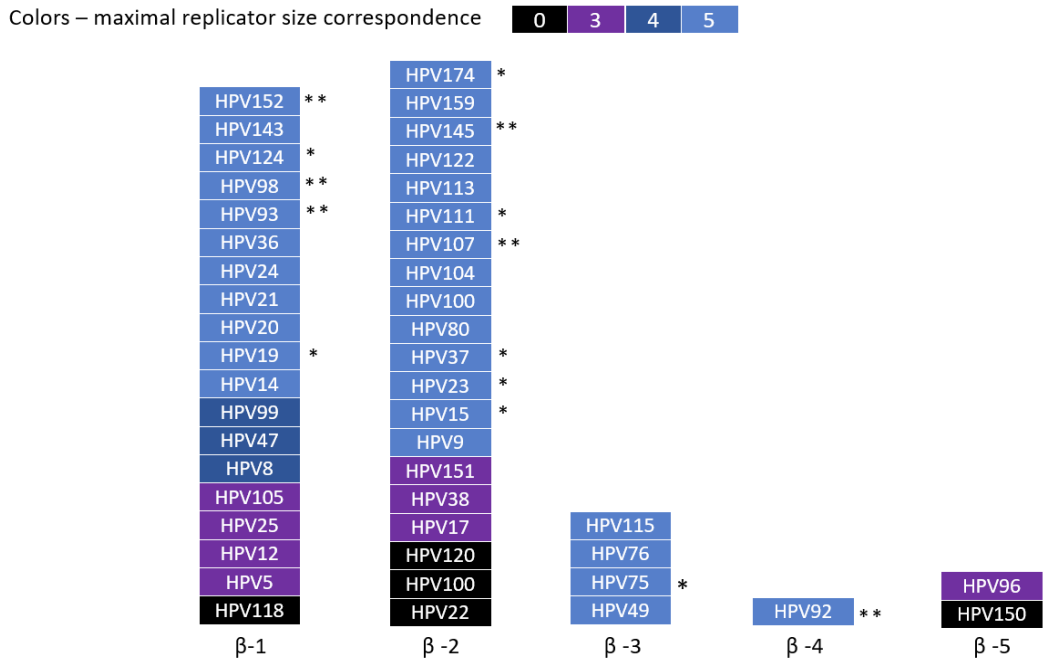

**Fig. 6.** The maximum size of replicators for different species and types of genus *β* human papillomavirus. The sets of many replicator patterns are non-monotonic: one (one asterisk) or two (two asterisks) replicators of length less than the maximum do not exist. Pay attention to a small part of virus types for which there are no replicators (black boxes).

Genus: *Gammapapillomavirus*

*γ*- papillomavirus genus is highly diverse, but most healthy adults chronically shed *γ*-virions from apparently healthy skin surfaces. Recent metagenomic studies have nearly doubled the number of known *γ*- HPV types [33]. While the *β*-papillomavirus genus is related to epidermodysplasia verruciforma, patients with the WHIM syndrome (warts, hypogammaglobulinemia, infections, myelokathexis) have been found to be uniquely susceptible to *γ* HPV-associated skin warts. NRA of *γ*-papillomaviruses shows that they share some properties with *β*-papillomaviruses, but also differ from them. Like *β*-papillomaviruses, their types can form replicators with a maximum length of up to 5. More importantly, the only kind of non-periodic transmission pattern is the same as that of *β*-papillomaviruses. The number of species of *γ*-papillomaviruses is large and, as can be seen from Fig. 7, the proportion of *γ*-papillomaviruses that do not generate neural replicators (black boxes) exceeds 60%, while for *β*-papillomaviruses this figure is about 11%. Thus, we can assume that *α*-14 papillomavirus species are more similar to *β*-, and not to *γ*-human papillomaviruses.

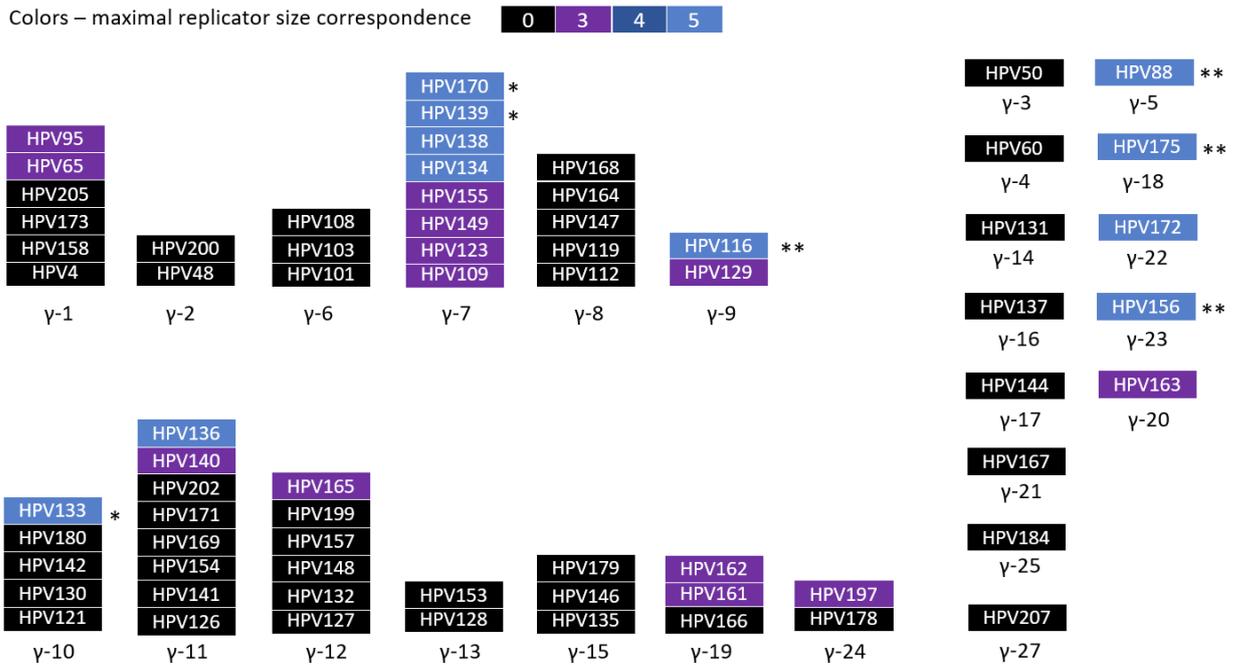

**Fig. 7.** The maximal size of replicators for different species and types of genus γ of human papillomaviruses. The sets of some replicator patterns are non-monotonic, with one (one asterisk) or two (two asterisks) less than maximum replicators missing. Pay attention to most of the types of viruses for which there are no replicators (black boxes).

Genus: *Mupapillomavirus*

*μ*-papillomaviruses are among the HPV types associated with cutaneous disease. The HPV1 type is responsible for about 30% cases of common warts [34]. The results obtained for *μ* human papillomaviruses show that they are similar to those for *β*- and *γ*-papillomaviruses (Fig.8 - left): the type HPV63 has the same non-periodic transmitted pattern as for the *β*- and *γ*-genera.

Genus: *Nupapillomavirus*

The most interesting result of NRA was obtained for *ν* (HPV41) human papillomavirus. Initially, this virus was isolated from a facial wart, but subsequently its DNA was found in some skin carcinomas and precancerous keratoses [35]. The genomic sequence of this virus is most distantly related to all other types of human papillomaviruses, and HPV41 virus has been identified as the first type of new genus *ν*. But NRA analysis shows that it is ideal for *α*-2 species because it has a maximal replicator size equals to 7 as well as the same periodic transmitted pattern (Fig. 8 - right). The clinical manifestations of HPV41 infection are similar to those of the types of *α*-2 species (although it also causes malignant skin lesions), so this result is not inconsistent with the characteristics of this genus. What also interesting is that NRA may provide some additional information about the problem of virus transfer to another host, as well as the taxonomy of viruses. As the Van Doorslayer paper says [36]:

> "*Because of the absence of cross-species infections, it is unlikely that horizontal gene transfer played any role in the evolution of the Papillomaviridae. In fact, a study specifically looking at the influence of horizontal gene transfer identified only a single potential cross-species transmission event. This event involved ancestors of a porcupine (EdPV1) and human (HPV41) papillomavirus* [37]. *These two viruses are the only members of a divergent genus (Nu papillomavirus); it will be of interest to see how the inclusion of more viruses in this genus will affect the conclusion of cross-species infection*".

In this situation, it was very interesting to use NRA to study the porcupine EdPV1 virus. It turned out that indeed it has a replicator of maximum size 6 (7 for HPV41), but the pattern transmitted by this replicator (3-periodic) differs significantly not only from the 2-period pattern of HPV41, but also from any pattern transmitted by the replicators of all papillomaviruses (Fig. 8). Thus, from the point of view of the NRA, the porcupine $\sigma$-virus EdPV1 cannot be combined with the HPV41 virus into one genus, nor can it be attached to any other genera of human papillomaviruses.

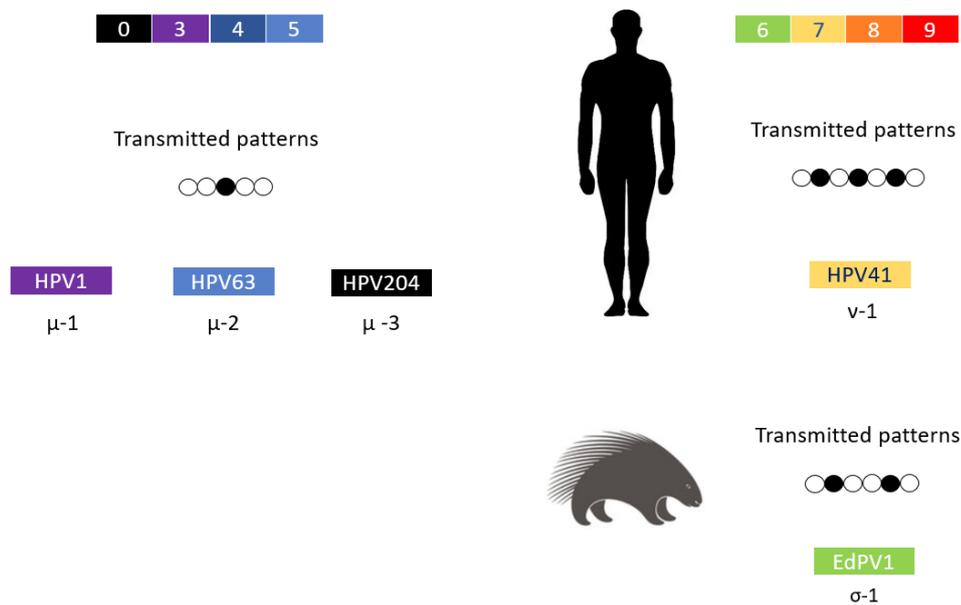

**Fig. 8.** Maximum size of replicators for different species and types of $\mu$-genus (upper left) and $\nu$-genus (upper right) of human papillomavirus and porcupine papillomavirus EdPV1 (lower right). The patterns transmitted by replicators are presented. The pattern of the EdPv1 virus differs both from the 2-period pattern of HPV41 (and all viruses of $\alpha$- genus) and from the aperiodic pattern of viruses belonging to $\mu$-, $\beta$- and $\gamma$- genera.

Now we can start to form a table which cells are defined by the motifs of replicators generated using WS- and KM-encoded genomes of human papillomaviruses (Fig. 9)

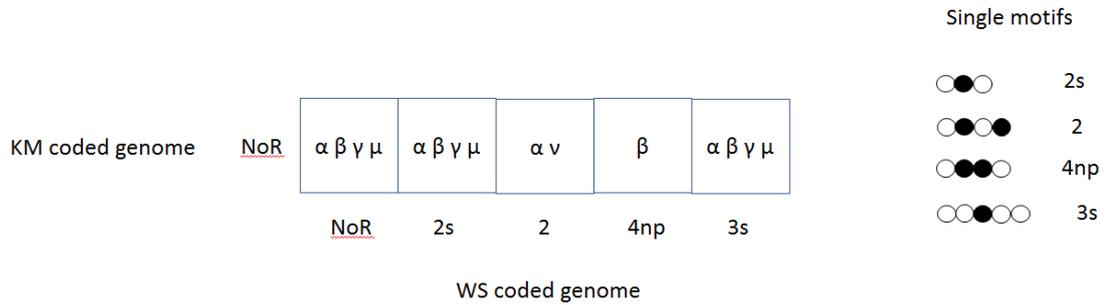

**Fig. 9.** According to the NRA, human papillomavirus genomes of genera $\alpha$, $\beta$, $\gamma$, $\mu$ and $\nu$ can be placed in the cells of the row or of the table of the size 1x5 (left side), where different columns correspond to the forms of single replicator motifs obtained using WS-encoded genomes (right side), while raw index NoR corresponds to the absence of replicators for their WS- and KM-encoded genomes. The notation 2s and 3s are used both to denote symmetrical patterns, which can also be considered as the seeds of 2-period pattern (3/2 $T$) and also 3-period pattern (5/3$T$). The notation 4np is used to denote non periodic patterns of the length $K$=4, "2" is used to denote all 2-period patterns.

This table can be expanded when considering other families of viruses. One of them, also associated with the occurrence of various tumors in humans, as well as great apes, other animals and birds is the family *Polyomaviridae*.

## 3. Neural Replicator Analysis of polyomaviruses

More variants of table cells occupation can be revealed by studying the viruses belonging to the *Polyomaviridae* family. The family *Polyomoviridae* contains tumor-causing viruses that infect mammals and birds. It includes 6 genera: *Alhapolyomavirus* (51 species), *Betapolyomavirus* (41 species), *Gammapolyomavirus* (9 species), *Deltapolyomavirus* (7 species), *Epsilonpolyomavirus* (3 species) and *Zetapolyomavirus* (1 species) that contain in overall 112 species. Virus genomes are circular double stranded DNA containing from 4776 to 5431 bp. Reconstructed evolutionary relationships are used for the delineation of genera and are derived from analyses of LTAg amino acid sequences [38].

Species demarcation criteria are as follows:

1. Sufficient information on the natural host.
2. Observed genetic distance from a member of the most closely related species is >15% nucleotide difference for the LTAg coding sequence.
3. When two polyomaviruses exhibit <15% observed genetic distance (as defined above), biological properties may be of additional critical importance (e.g. host specificity, disease association, tissue tropism, etc.)

Genus: *Alphapolyomavirus*

Many genomes of viruses of the genus *Alhapolyomavirus* belong to the "*black hole*" cell (NoR, NoR) which, as we saw earlier, contains cancer inducing types of human papillomavirus. This is also true for *Alphapolyomavirus*. Such dangerous human virus as *Merkel cell carcinoma* also lacks replicators for the WS- and KM-encoded genomes. This cell is also occupied by virus genomes of some great apes, monkeys, bats and some other animals (Table 3.1).

Table 3.1. Virus genomes in a cell (NoR, NoR).

| № | Species | Virus name | Abbreviation | Accession |
|---|---|---|---|---|
| 1 | *Alphapolyomavirus quitihominis* | Merkel cell carcinoma virus | MCPyV | HM011556 |
| 2 | *Alphapolyomavirus panos* | Chimpanzee polyomavirus | ChPyV | FR692334 |
| 3 | *Alphapolyomavirus quintipanos* | Pan troglodytes verus polyomavirus 4 | PtrovPyV4 | JX159981 |
| 4 | *Alphapolyomavirus sextipanos* | Pan troglodytes verus polyomavirus 5 | PtrovPyV5 | JX159982 |
| 5 | *Alphapolyomavirus septipanos* | Pan troglodytes schweinfurthii polyomavirus 2 | PtrosPyV2 | JX159983 |
| 6 | *Alphapolyomavirus ponabelii* | Sumatran orang-utan polyomavirus | OraPyV-Sum | FN356901 |
| 7 | *Alphapolyomavirus macacae* | Macaca fascicularis polyomavirus 1 | JX159986 | MfasPyV1 |
| 8 | *Alphapolyomavirus pirufomitratus* | Piliocolobus rufomitratus polyomavirus 1 | JX159984 | PrufPyV1 |
| 9 | *Alphapolyomavirus chlopygerythrus* | vervet monkey polyomavirus 1 | AB767298 | VmPyV1 |
| 10 | *Alphapolyomavirus eidola* | Eidolon polyomavirus 1 | JX520660 | EidolonPyV |
| 11 | *Alphapolyomavirus dobsoniae* | bat polyomavirus 5a | AB972945 | BatPyV5a |
| 12 | *Alphapolyomavirus sominutus* | Sorex minutus polyomavirus 1 | MF401583 | ScorPyV1 |

The nearest cell (2s, NoR), characterized by the presence of the shortest simple replicator (1 -1 1) generated by neural network having $K=3$ neurons and processing WS-encoded genome is also populated by polyomaviruses of human, great apes and monkeys but differs from cell (NoR, NoR) by the presence of many bat polyomaviruses (Table 3.2).

Table 3.2. Virus genomes in a cell (2s, NoR).

| № | Species | Virus name | Abbreviation | Accession |
|---|---|---|---|---|
| 1 | *Alphapolyomavirus terdecihominis* | New Jersey polyomavirus | NJPyV | KF954417 |
| 2 | *Alphapolyomavirus quardecihominis* | LI polyomavirus | LIPyV | KY404016 |
| 3 | *Alphapolyomavirus octihominis* | Trichodysplasia spinulosa-associated polyomav. | TSPyV | GU989205 |
| 4 | *Alphapolyomavirus cardiodermae* | Cardioderma polyomavirus | CardiodermaPyV | JX520659 |
| 5 | *Alphapolyomavirus secupanos* | Pan troglodytes verus polyomavirus 1a | PtrovPyV1a | HQ385746 |
| 6 | *Alphapolyomavirus tertipanos* | Pan troglodytes verus polyomavirus 2a | PtrovPyV2a | HQ385748 |
| 7 | *Alphapolyomavirus quartipanos* | Pan troglodytes verus polyomavirus 3 | PtrovPyV3 | JX159980 |
| 8 | *Alphapolyomavirus ponpygmaeus* | Bornean orang-utan polyomavirus | OraPyV-Bor | FN356900 |
| 9 | *Alphapolyomavirus gorillae* | Gorilla gorilla gorilla polyomavirus 1 | GgorgPyV1 | HQ385752 |
| 10 | *Alphapolyomavirus pibadius* | Piliocolobus badius polyomavirus 2 | PbadPyV2 | KX509984 |
| 11 | *Alphapolyomavirus secarplanirostris* | bat polyomavirus 3a-A1055 | BatPyV3a-A1055 | JQ958886 |
| 12 | *Alphapolyomavirus sturnirae* | bat polyomavirus 3a-B0454 | BatPyV3a-B0454 | JQ958888 |
| 13 | *Alphapolyomavirus omartiensseni* | Otomops polyomavirus 1 | OtomopsPyV1 | JX520664 |
| 14 | *Alphapolyomavirus molossi* | bat polyomavirus 3b | BatPyV3b | JQ958893 |
| 15 | *Alphapolyomavirus carolliae* | bat polyomavirus 4b | BatPyV4b | JQ958889 |
| 16 | *Alphapolyomavirus acelebensis* | bat polyomavirus 5b2 | BatPyV5b-2 | AB972940 |
| 17 | Alphapolyomavirus *saraneus* | Sorex araneus polyomavirus 1 | SaraPyV1 | MF374997 |
| 18 | *Alphapolyomavirus socoronatus* | Sorex coronatus polyomavirus 1 | SminPyV1 | MF374999 |
| 19 | *Alphapolyomavirus procyonis* | raccoon polyomavirus | RacPyV | JQ178241 |
| 20 | *Alphapolyomavirus philantombae* | Philantomba monticola polyomavirus 1 | PmonPyV1 | MG654482 |

The cell (2, NoR), which is very informative for papillomaviruses (recall that it contains the genomes of skin wart viruses, and these viruses demonstrate 2-periodicity of single motifs of WS-encoded genomes), contains only one bat *α*-polyomavirus with a maximum replicator size of *K*=8 (Table 3.3).

Table 3.3. Virus genomes in a cell (2, NoR).

| № | Species | Virus name | Abbreviation | Accession |
|---|---|---|---|---|
| 1 | *Alphapolyomavirus tertarplanisrostris* | bat polyomavirus 4a | BatPyV4a | JQ958890 |

The next cell (3s, NoR) which usually contains the majority of human *β*-papillomaviruses having single motif (1 1 – 1 1 1) for WS-encoded genomes, also contains many *α*-polyomaviruses, with the exception of the great ape viruses (Table 3.4).

Table 3.4. Virus genomes in a cell (3s, NoR).

| № | Species | Virus name | Abbreviation | Accession |
|---|---|---|---|---|
| 1 | *Alphapolyomavirus nonihominis* | human polyomavirus 9 | HPyV9 | HQ696595 |
| 2 | *Alphapolyomavirus tertichlopygerythrus* | vervet monkey polyomavirus 3 | VmPyV3 | AB767297 |
| 3 | *Alphapolyomavirus pacynocephalus* | yellow baboon polyomavirus 1 | YbPyV1 | AB767294 |
| 4 | *Alphapolyomavirus apaniscus* | Ateles paniscus polyomavirus 1 | ApanPyV1 | JX159987 |
| 5 | *Alphapolyomavirus mischreibersii* | Miniopterus schreibersii polyomavirus 1 | MschPyV1 | LC185213 |
| 6 | *Alphapolyomavirus secumischreibersii* | Miniopterus schreibersii polyomavirus 2 | MschPyV2 | LC185216 |
| 7 | *Alphapolyomavirus secomartiensseni* | Otomops polyomavirus 2 | OtomopsPyV2 | JX520658 |
| 8 | *Alphapolyomavirus ptevampyrus* | bat polyomavirus 5b1 | BatPyV5b-1 | AB972944 |
| 9 | *Alphapolyomavirus tuglis* | Tupaia glis polyomavirus 1 | TgliPyV1 | MG721015 |
| 10 | *Alphapolyomavirus tubelangeri* | Tupaia belangeri polyomavirus | | MK443498 |
| 11 | *Alphapolyomavirus callosciuri* | Callosciurus erythraeus polyomavirus 1 | CeryPyV1 | MK671087 |

All the *α*-polyomaviruses considered above are located in cells already occupied by human papillomaviruses. Now consider those polyomaviruses that occupy new cells. The hosts of these viruses are presumably rodents and mice. The replicators of some of them have obvious mixtures of 2- and 3-periodic motifs, or only pure 3-periodic motifs for WS-encoded genomes. Accordingly, two new cells are introduced: (2-3, NoR) and (3, NoR) − Tables 3.5 and 3.6.

Table 3.5. Virus genomes in a cell (2-3, NoR).

| № | Species | Virus name | Abbreviation | Accession |
|---|---|---|---|---|
| 1 | *Alphapolyomavirus ranorvegicus* | rattus norvegicus polyomavirus 1 | RnorPyV1 | KR075943 |
| 2 | *Alphapolyomavirus muris* | mouse polyomavirus | MPyV | AF442959 |

Table 3.6. Virus genomes in a cell (3, NoR).

| № | Species | Virus name | Abbreviation | Accession |
|---|---|---|---|---|
| 1 | *Alphapolyomavirus aflavicollis* | apodemus flavicollis polyomavirus 1 | AflaPyV1 | MG654476 |
| 2 | *Alphapolyomavirus mauratus* | hamster polyomavirus | HaPyV | JX036360 |
| 3 | *Alphapolyomavirus secumastomysis* | mastomys natalensis polyomavirus 2 | MnatPyV2 | MG701350 |
| 4 | *Alphapolyomavirus tertimastomysis* | mastomys natalensis polyomavirus 3 | MnatPyV3 | MN417229 |

But more interestingly, among *α*-polyomaviruses, we encounter a virus that has replicator motifs with 2-periodicity (for $K=8$) and 3-periodicity (for $K$ up to 23) for WS-encoded genomes and, for the first time, motif with 3-periodicity for KM-encoded genome (for $K=26$)! The corresponding cell can be named as (2-3, 3) – Table 3.7.

Table 3.7. Virus genomes in a cell (2-3, 3).

| № | Species | Virus name | Abbreviation | Accession |
|---|---|---|---|---|
| 1 | *Alphapolyomavirus suis* | sus scrofa polyomavirus 1 | SscrPyV1 | KR065722 |

Genus: *Betapolyomavirus*

The genomes of two viruses (BKPyV and JCPyV) associated with dangerous human diseases that occurin immune deficit patients−nephropathy and progressive multifocal leukoencephalopathy, correspondingly − occupy the *"black hole"* cell (NoR, NoR) – Table 3.8.

Table 3.8. Virus genomes in a cell (NoR, NoR).

| № | Species | Virus name | Abbreviation | Accession |
|---|---|---|---|---|
| 1 | *Betapolyomavirus hominis* | BK polyomavirus | BKPyV | V01108 |
| 2 | *Betapolyomavirus secuhominis* | JC polyomavirus | JCPyV | J02226 |
| 3 | *Betapolyomavirus quartihominis* | WU polyomavirus | WUPyV | EF444549 |
| 4 | *Betapolyomavirus octipanos* | pan troglodytes verus polyomavirus 8 | PtrovPyV8 | KT884050 |
| 5 | *Betapolyomavirus calbifrons* | cebus albifrons polyomavirus 1 | CalbPyV1 | JX159988 |
| 6 | *Betapolyomavirus canis* | canis familiaris polyomavirus 1 | CfamPyV1 | KY341899 |
| 7 | *Betapolyomavirus desrotundus* | bat polyomavirus 2a | BatPyV2a | JQ958892 |
| 8 | *Betapolyomavirus pteparnellii* | bat polyomavirus 2b | BatPyV2b | JQ958891 |
| 9 | *Betapolyomavirus secudobsoniae* | bat polyomavirus 6b | BatPyV6b | AB972947 |
| 10 | *Betapolyomavirus pantherae* | panthera leo polyomavirus 1 | PleoPyV1 | MG701353 |
| 11 | *Betapolyomavirus vicugnae* | alpaca polyomavirus | AlPyV | KU879245 |
| 12 | *Betapolyomavirus tertimuris* | mus musculus polyomavirus 3 | MPoV3 | MF175082 |

The nearest cell (2s, NoR) corresponding to the shortest simple replicator (1 -1 1) is populated by polyomaviruses listed in Table 3.9. There are no human and great ape viruses here, but there is significant amount of bat viruses, as well as two seal viruses.

Table 3.9. Virus genomes in a cell (2s, NoR).

| № | Species | Virus name | Abbreviation | Accession |
|---|---|---|---|---|
| 1 | *Betapolyomavirus macacae* | simian virus 40 | SV40 | J02400 |
| 2 | *Betapolyomavirus saboliviensis* | squirrel monkey polyomavirus | SquiPyV | NC_009951 |
| 3 | *Betapolyomavirus mafricanus* | miniopterus polyomavirus | MiniopterusPyV | JX520661 |
| 4 | *Betapolyomavirus myolucifugus* | myotis polyomavirus | MyPyV | FJ188392 |
| 5 | *Betapolyomavirus raegyptiacus* | rousettus aegyptiacus polyomavirus 1 | RaegPyV1 | LC185218 |
| 6 | *Betapolyomavirus arplanirostris* | bat polyomavirus 2c | BatPyV2c | JQ958887 |
| 7 | *Betapolyomavirus secacelebensis* | bat polyomavirus 6a | BatPyV6a | AB972941 |
| 8 | *Betapolyomavirus tertidobsoniae* | bat polyomavirus 6c | BatPyV6c | AB972946 |
| 9 | *Betapolyomavirus mastomysis* | mastomys polyomavirus | MasPyV | AB588640 |
| 10 | *Betapolyomavirus lepweddellii* | weddell seal polyomavirus | WsPyV | KX533457 |
| 11 | *Betapolyomavirus zacalifornianus* | california sea lion polyomavirus 1 | SLPyV | GQ331138 |

Unlike *α*-polyomaviruses, *β*-polyomaviruses generate many replicators with 2-periodic motifs, thus populating the cell (2, NoR) (Table 3.10).

Table 3.10 Virus genomes in a cell (2, NoR).

| № | Species | Virus name | Abbreviation | Accession |
|---|---|---|---|---|
| 1 | *Betapolyomavirus callosciuri* | callosciurus prevostii polyomavirus 1 | CprePyV1 | MK883808 |
| 2 | *Betapolyomavirus equi* | equine polyomavirus | EPyV | JQ412134 |
| 3 | *Betapolyomavirus gliris* | glis glis polyomavirus 1 | GgliPyV1 | MG701352 |
| 4 | *Betapolyomavirus marvalis* | microtus arvalis polyomavirus 1 | CVPyV | KR612373 |
| 5 | *Betapolyomavirus secuchlopygerythrus* | vervet monkey polyomavirus 2 | VmPyV2 | AB767299 |

Of particular interest is the case of microtus arvalis polyomavirus 1 (its RT is shown in Fig. 10).

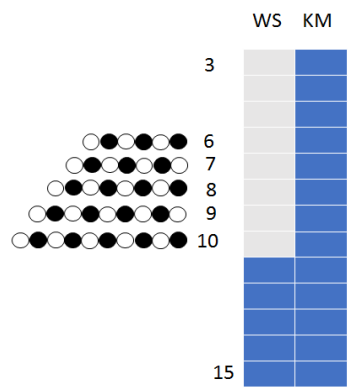

**Fig. 10.** RT of microtus arvalis polyomavirus genome 1 shows the existence of many simple single 2-periodic pattern replicators of sizes up to $K=10$ for the case of the WS-encoded genome.

When considering human papillomaviruses, we already found relatively rare cases of virus genomes for which maximum replicator size was 4 but corresponding motif is not 2-periodic (Fig. 9). Such a case is also found for meles meles polyomavirus 1 and its genome can be placed in cell (4, np) (Table 3.11).

Table 3.11. Virus genomes in a cell (4, np).

| № | Species | Virus name | Abbreviation | Accession |
|---|---|---|---|---|
| 1 | *Betapolyomavirus meletis* | meles meles polyomavirus 1 | MmelPyV1 | KP644238 |

A cell (3s, NoR) that corresponds to genome replicators that has at least one single WS-encoded motif (1 1 -1 1 1) (for some viruses replicator exchanges two patterns) also contains many members of the genus $\beta$ polyomaviruses (Table 3.12).

Table 3.12. Virus genomes in a cell (3s, NoR).

| № | Species | Virus name | Abbreviation | Accession |
|---|---|---|---|---|
| 1 | *Betapolyomavirus tertihominis* | KI polyomavirus | KIPyV | EF127906 |
| 2 | *Betapolyomavirus secupacynocephalus* | yellow baboon polyomavirus 2 | YbPyV2 | AB767295 |
| 3 | *Betapolyomavirus sasciureus* | saimiri sciureus polyomavirus 1 | SsciPyV1 | JX159989 |
| 4 | *Betapolyomavirus ptedavyi* | pteronotus polyomavirus | PteronotusPyV | JX520662 |
| 5 | *Betapolyomavirus cercopitheci* | cercopithecus erythrotis polyomavirus 1 | CeryPyV1 | JX159985 |
| 6 | *Betapolyomavirus sciuri* | sciurus carolinensis polyomavirus 1 | ScarPyV1 | MK671101 |
| 7 | *Betapolyomavirus elephanti* | African elephant polyomavirus 1 | AelPyV1 | KF147833 |
| 8 | *Betapolyomavirus secumuris* | mouse pneumotropic virus | MPtV | KT987216 |
| 9 | *Betapolyomavirus myoglareolus* | myodes glareolus polyomavirus 1 | BVPyV | KR612368 |

As with $\alpha$-polyomaviruses, some $\beta$-polyomaviruses have 3-period motifs for WS-encoded genomes and must be placed in a cell (3, NoR) –Table 3.13. Note that, as in the case of $\alpha$-polyomaviruses, this cell includes rat polyomavirus and also polyomavirus of hare, and that hare feeds on tree bark. This fact will be further considered as valuable in the analysis of plant viruses.

Table 3.13. Virus genomes in a cell (3, NoR).

| № | Species | Virus name | Abbreviation | Accession |
|---|---|---|---|---|
| 1 | *Betapolyomavirus enhydrae* | sea otter polyomavirus |  | KM282376 |
| 2 | *Betapolyomavirus leporis* | lepus polyomavirus 1 | LPyV1 | MN994868 |
| 3 | *Betapolyomavirus securanorvegicus* | rat polyomavirus 2 | RatPyV2 | KX574453 |

Genus: *Gammapolyomavirus*

*Gammapolyomavirus* is an extremely interesting genus for the NRA. Their avian viruses colonize only cells corresponding to 2-periodic (one virus) and 3-periodic WS motifs (5 viruses) and also one virus with 3-periodic WS-encoded motif and 5-periodic KM-encoded motifs (Tables 3.14 and 3.15).

Table 3.14. Virus genomes in a cell (2, NoR).

| № | Species | Virus name | Abbreviation | Accession |
|---|---|---|---|---|
| 1 | *Gammapolyomavirus anseris* | goose hemorrhagic polyomavirus | GHPV | AY140894 |

Table 3.15. Virus genomes in a cell (3, NoR).

| № | Species | Virus name | Abbreviation | Accession |
|---|---|---|---|---|
| 1 | *Gammapolyomavirus corvi* | crow polyomavirus | CpyV | DQ192570 |
| 2 | *Gammapolyomavirus cratorquatus* | butcherbird polyomavirus | Butcherbird PyV | KF360862 |
| 3 | *Gammapolyomavirus padeliae* | Adélie penguin polyomaviru | ADPyV | KP033140 |
| 4 | *Gammapolyomavirus pypyrrhula* | finch polyomavirus | FpyV | DQ192571 |
| 5 | *Gammapolyomavirus secanaria* | canary polyomavirus | CaPyV | GU345044 |

Budgerigar fledgling disease virus can be placed in a new cell (3,5) unconditionally, while erythrura gouldiae polyomavirus only conditionally (Fig. 11 and Table 3.16)

Table 3.16. Virus genomes in a cell (3, 5).

| № | Species | Virus name | Abbreviation | Accession |
|---|---|---|---|---|
| 1 | *Gammapolyomavirus avis* | budgerigar fledgling disease virus | BFDV | AF241168 |
| 2 | *Gammapolyomavirus egouldiae* | Erythrura gouldiae polyomavirus 1 | EgouPyV1 | KT302407 |

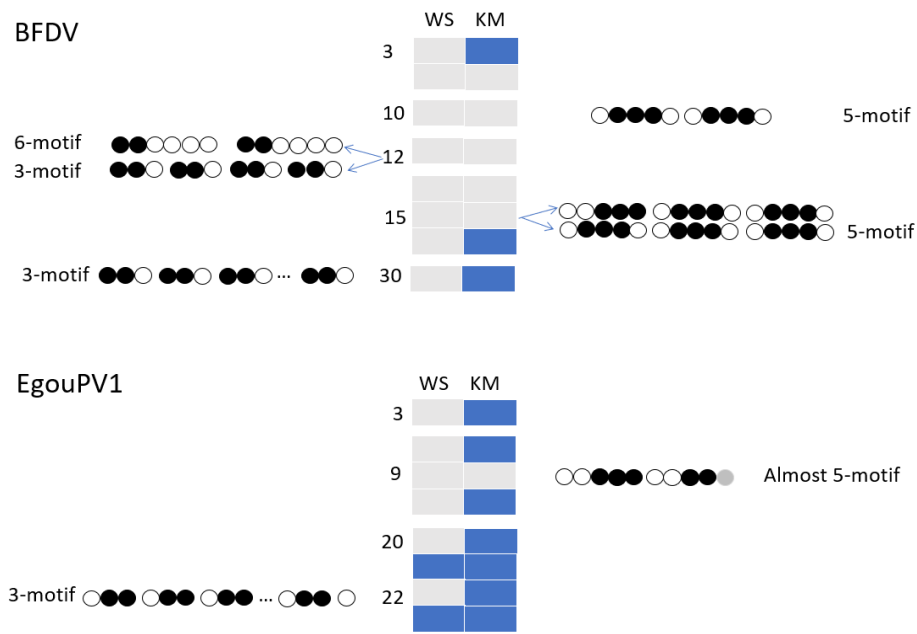

**Fig. 11.** RTs of the budgerigar fledgling disease virus (BFDV) genome and the Erythrura gouldiae (EgouPV1) polyomavirus 1 genome. Note, that for the WS-encoded genome of BFDV, 6-periodic motif appears as one transmitted pattern of complex replicator of the size $K$=12, but then only 3-periodicity is retained. The presence of the 5-periodicity is evident for KM-encoded genome of BFDV, while for EgouPV1 the second period of the 5-periodical sequence is distorted by one bit.

The last Hungarian finch polyomavirus has for the first glance hardly interpretable motifs for KM-encoded genome (in section 5 we will encounter a very similar motif for the replicator of WS-encoded genome of Mirabilis mosaic virus).

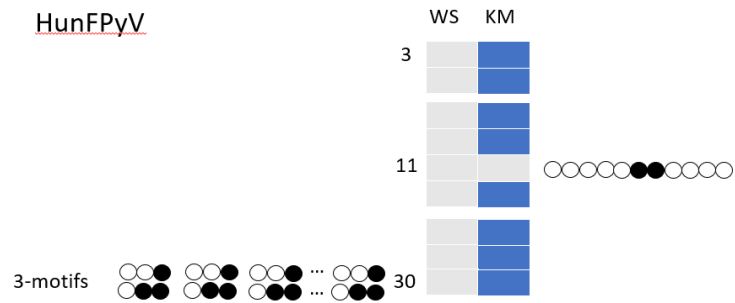

**Fig. 12.** RT of Hungarian finch polyomavirus genome. The unique replicator of size K=11 for KM-encoded genome is nearly symmetrical but hardly classified.

Genus: *Deltapolyomavirus*

This genus contains seven species. It is remarkable that its members such as human polyomavirus 6 (HPyV6) and human polyomavirus 7 (HPyV7), found on the skin, are reasonably placed in a cell (3s, NoR), populated as we saw earlier, with human papillomaviruses of genus *β*, which also found on the skin of immunosuppressed patients. The remaining members detected in gastrointestinal tract, are distributed over cells (NoR, NoR), (2s, NoR), (3s, NoR), (3, NoR) – Tables 3.17–3.20.

Note, that the genome of the raccoon virus (PlotPyV1) is placed in a cell (3, NoR) inhabited by rodent polyomaviruses and also lepus polyomavirus. Like these mammals, raccoon can also find his food in trees.

Table 3.17. Virus genomes in a cell (NoR, NoR).

| № | Species | Virus name | Abbreviation | Accession |
|---|---|---|---|---|
| 1 | *Deltapolyomavirus decihominis* | MW polyomavirus | MWPyV | JQ898291 |

Table 3.18. Viruses whose genomes enter a cell (2s, NoR).

| № | Species | Virus name | Abbreviation | Accession |
|---|---|---|---|---|
| 1 | *Deltapolyomavirus undecihominis* | STL polyomavirus | STLPyV | JX463183 |

Table 3.19. Virus genomes in a cell (3s, NoR).

| № | Species | Virus name | Abbreviation | Accession |
|---|---|---|---|---|
| 1 | *Deltapolyomavirus sextihominis* | human polyomavirus 6 | HPyV6 | HM011560 |
| 2 | *Deltapolyomavirus septihominis* | human polyomavirus 7 | HPyV7 | HM011566 |
| 3 | *Deltapolyomavirus ailuropodae* | giant panda polyomavirus | AmelPyV1 | KY612371 |
| 4 | *Deltapolyomavirus canis* | Canis lupus polyomavirus 1 | ClupPyV1 | MG701355 |

Table 3.20. Virus genomes in a cell (3, NoR).

| № | Species | Virus name | Abbreviation | Accession |
|---|---|---|---|---|
| 1 | *Deltapolyomavirus secuprocyonis* | raccoon-associated polyomavirus 2 | PlotPyV1 | KY549442 |

Genus: *Epsilonpolyomavirus*

The members of this genus that infect cetartiodactyls, are distributed over cells (NoR, NoR), (2s, NoR) and (3, NoR) –Tables 3.21- 3.23. Once again, we note that the host of the virus, which occupies the last cell, the potamochoerus or river pig, also feeds on plants.

Table 3.21. Virus genomes in a cell (NoR, NoR).

| № | Species | Virus name | Abbreviation | Accession |
|---|---|---|---|---|
| 1 | *Epsilonpolyomavirus caprae* | Capra aegragus polyomavirus 1 | CaegPyV1 | MG654479 |

Table 3.22. Virus genomes in a cell (2s, NoR).

| № | Species | Virus name | Abbreviation | Accession |
|---|---|---|---|---|
| 1 | *Epsilonpolyomavirus bovis* | bovine polyomavirus | BPyV | D13942 |

Table 3.23. Virus genomes in a cell (3, NoR).

| № | Species | Virus name | Abbreviation | Accession |
|---|---|---|---|---|
| 1 | Epsilonpolyomavirus poporcus | Potamochoerus porcus polyomavirus 1 | PporPyV1 | MG654481 |

Genus: *Zetapolyomavirus*

The only member of this genus infects dolphins and occupies the cell (NoR, NoR).

Table 3.24. Virus genomes in a cell (NoR, NoR).

| № | Species | Virus name | Abbreviation | Accession |
|---|---|---|---|---|
| 1 | *Zetapolyomavirus delphini* | dolphin polyomavirus 1 | DPyV | KC594077 |

It can be concluded that polyomaviruses of six genera are placed both in the already mentioned cells occupied by human papillomaviruses and in new cells: (2-3, NoR), (2-3, 3), (3, NoR) and (3,5). So, a two-dimensional cellular structure begins to form (Fig.13). Further, it will be interesting to observe and discuss the location of genomes of viruses belonging to different families in the same cell.

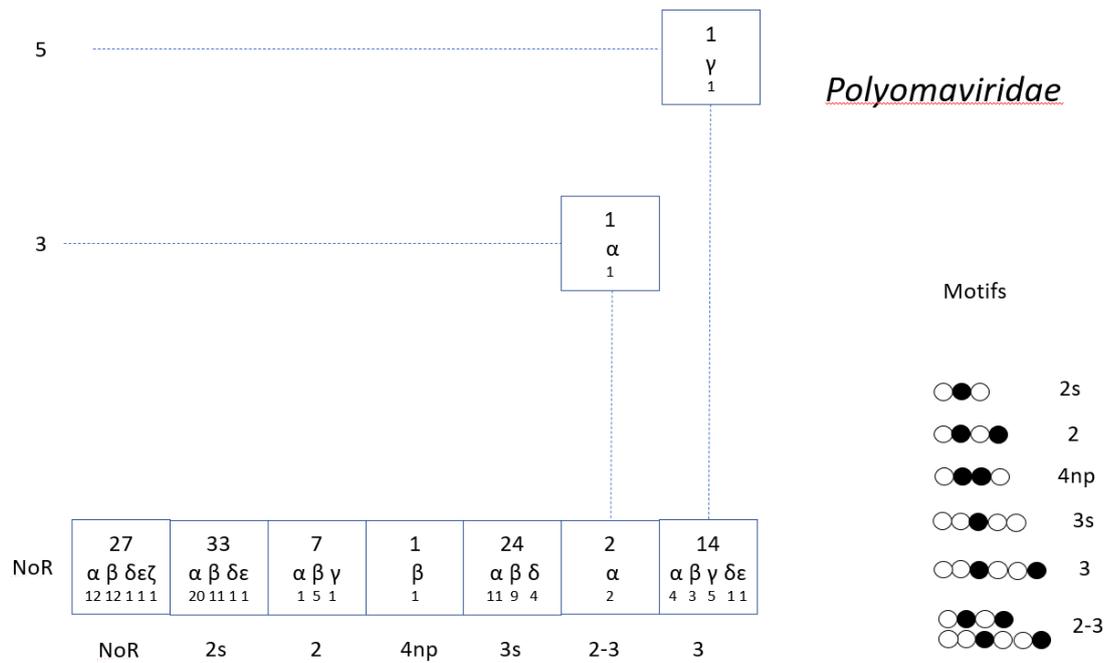

**Fig. 13.** Binomial table cells occupied by polyomavirus genomes (left). Each cell contains the following information: the total number of species, genera of species, and the number of species in each genus. Unlike papillomaviruses, not only the bottom row is filled (including two new cells (2-3, NoR) and (3, NoR) – typical patterns of replicators are shown on the right, but also two cells in the upper rows (2-3,3) and (3,5) where the SscrPyV1 and BFDV viruses are located (see Tables 3.7 and 3.16).

As an example, consider a cell (2, NoR) in which human papillomaviruses that cause warts are concentrated. This cell is also occupied by polyomaviruses which have as hosts various mammals, but not humans, great apes, birds and dolphins (Fig.14).

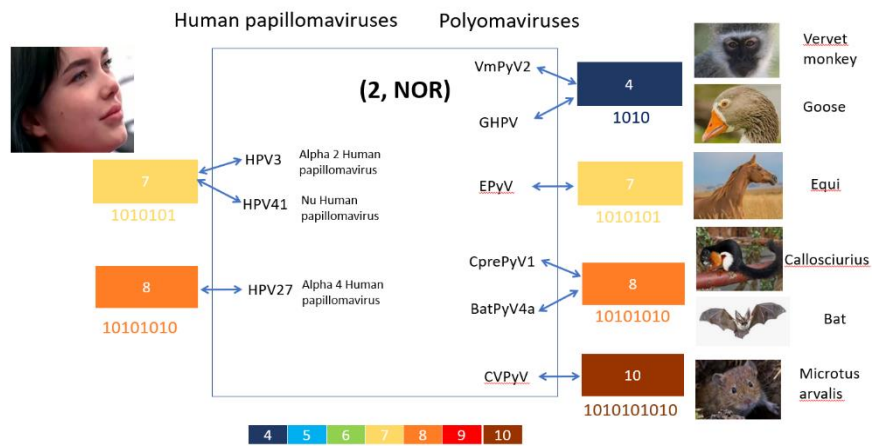

**Fig. 14.** Cell (2, NoR) is occupied by both *α*-2, *α*-4 and *ν* human papillomaviruses that cause common warts (shown at left) and have replicators with 2-periodic single patterns for WS-encoded genomes, as well as animal and avian polyomaviruses with similar replicators (shown on the right).

Another interesting observation can be made by considering the case of polyomaviruses which occupy the cell (3, NoR), but for this, NRA of plant viruses belonging to the family *Caulimoviridae* must first be performed.

## 4. Neural Replicator Analysis of the family *Caulimoviridae*

The family *Caulimoviridae* contains ten genera: *Badnavirus*, *Tungrovirus*, *Caulimovirus*, *Cavemovirus*, *Dioscovirus*, *Petuvirus*, *Rosadnavirus*, *Solendovirus, Soymovirus* and *Vaccinivirus*. Genus demarcation criteria do not contain genomic sequence similarity, but are based on virion morphology, number of ORF, host types (monocot or dicot) and transmission mode. In each genus, with the exception of those that have single member, demarcation of species is based in particular on a difference in polymerase (RT+RNAse H) nt sequences of more than 20%,

We start with the special genus of the family *Caulimoviridae* – *Badnavirus*, for which NRA provides interesting and important results.

Genus: *Badnavirus*

Badnaviruses are bacilliform non-enveloped double-stranded DNA pararetroviruses. Their genomes contain about 8kb of dsDNA with three to seven open reading frames (ORF). They are also one of the most important groups of plant viruses and have become serious pathogens affecting the cultivation of horticultural crops in the tropics. As noted in [39] the presence of endogenous badnaviruses poses a new challenge for reliable diagnostics and taxonomy.

NRA performed for various network sizes from $N$=3 to $N$=30 shows that:

1. Replicators appear for all WS-encoded bandavirus genomes.
2. The maximum length of replicator networks has different values up to 30 or more.
3. Sets of transmitted patterns have a different number of motifs.
4. With one exception the transmitted patterns for WS-encoded genomes are 3-periodic for at least sufficiently large values of $K$.
5. The number of patterns of networks of maximum length varies from 1 to 3.

A striking feature of the WS-encoded badnavirus genomes is the 3-periodicity of replicator motifs. The only exceptional case of Aglaonema bacilliform virus (ABV, NC_055236.1) can be considered as characterized by the presence of a pattern of 5/3 periods for $T$=3: ((1 1-1) 1 1), while all other members of this genus have transmitted patterns that have from 2 up to 10 or more periods. However, we will further consider ABV as a real exception also because the only virus of the genus Tungrovirus (RTBV, AF220561), which also has the form of a bacillus, does not have WS-encoded replicators at all. This exceptional virus, like many other badnaviruses, also has replicators for the KM-encoded genome and occupies the cell (3s, 3s) (Table 4.1).

Table 4.1. Virus genomes in a cell (3s, 3s).

| № | Species | Virus name | Abbreviation | Accession |
|---|---|---|---|---|
| 1 | *Aglaonema bacilliform virus* | Aglaonema bacilliform virus | ABV | MH384837 |

Further, with respect to papillomaviruses and many of polyomaviruses, there are a number of badnaviruses for which no replicators for KM-encoded genomes have emerged. Remarkably, except to the Polyscias mosaic virus, five badnaviruses which cause mosaic diseases (Cacao mild mosaic virus, Badnavirus castaneae, Citrus yellow mosaic virus, Jujube mosaic-associated virus, Pagoda yellow mosaic associated virus) belong to this class. Also 9 of 11 Cacao viruses belong to this cell. All badnaviruses of this kind obviously can be placed in cell (3, NoR) – Table 4.2. It is also interesting that many viruses from this cell have more than 3 open reading frames. It should also be noted, that 18 of 22 viruses have dicot hosts, many of which are trees (Birch, Bougainvillea, Cacao, Chestnut, Jujube, Lemon, Sichuan Pepper tree, Pagoda tree) and only 4 viruses have monocot hosts. It is remarkable, that it is this tree-reach cell is inhabited by polyomaviruses of mice, rats, hares and raccoons, for which such plants are food sources (Fig. 15).

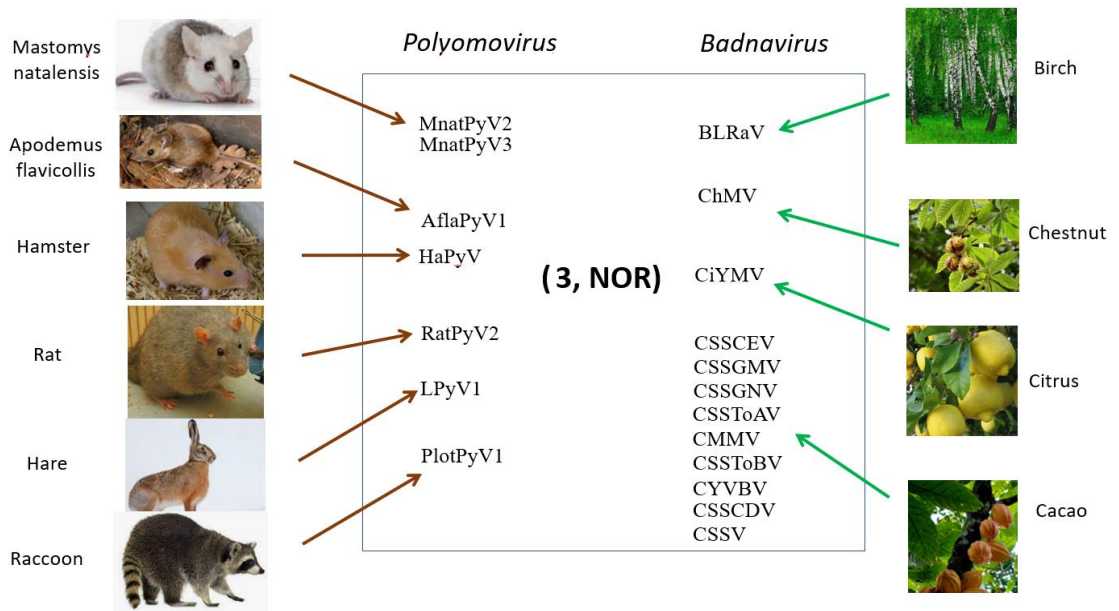

**Fig. 15.** The cell (3, NoR) characterized by the 3-periodicity of WS-encoded viral genomes and the absence of replicators for KM-encoded genomes, is inhabited by badnaviruses of trees (some of them are shown on the right) and, on the other hand, by polyomaviruses of animals that eat tree bark and fruits (they are shown on the left).

Table 4.2. Virus genomes in a cell (3, NoR).

| № | Species | Virus name | Abbreviation | Accession |
|---|---|---|---|---|
| 1 | *Banana streak UL virus* | banana streak UL virus | BSULV | NC_015504.1 |
| 2 | *Yacon necrotic mottle virus* | Yacon necrotic mottle virus | YNMoV | NC_026472.1 |
| 3 | *Dioscorea bacilliform AL virus 2* | Dioscorea bacilliform AL virus 2 | DBALV 2 | MH404164.1 |
| 4 | *Taro bacilliform CH virus* | Taro bacilliform CH virus | TaBCHV | MG833014.1 |
| 5 | *Bougainvillea chlorotic vein banding virus* | Bougainvillea chlorotic vein banding virus | BCVBV | MK816926 |
| 6 | *Cacao swollen shoot CE virus* | Cacao swollen shoot CE virus | CSSCEV | NC_040692.1 |
| 7 | *Badnavirus castaneae* | Chestnut mosaic virus | ChMV | MT269853.1 |
| 8 | *Cacao swollen shoot Ghana M virus* | Cacao swollen shoot Ghana M virus | CSSGMV | NC_043534 |
| 9 | *Cacao swollen shoot Ghana N virus* | Cacao swollen shoot Ghana N virus | CSSGNV | NC_040622.1 |
| 10 | *Cacao swollen shoot Togo A virus* | Cacao swollen shoot Togo A virus | CSSToAV | MF642716 |
| 11 | *Wisteria badnavirus 1* | Wisteria badnavirus 1 | WBV1 | NC_034252.1 |
| 12 | *Jujube mosaic-associated virus* | Jujube mosaic-associated virus | JuMaV | NC_035472 |
| 13 | *Cacao mild mosaic virus* | Cacao mild mosaic virus | CMMV | NC_033738.1 |
| 14 | *Citrus yellow mosaic virus* | Citrus yellow mosaic virus | CiYMV | NC_003382 |
| 15 | *Canna yellow mottle-associated virus* | Canna yellow mottle-associated virus | CaYMV | NC_030462 |
| 16 | *Pagoda yellow mosaic associated virus* | Pagoda yellow mosaic associated virus | PYMaV | NC_024301 |
| 17 | *Birch leaf roll-associated virus* | Birch leaf roll-associated virus | BLRaV | MG686419.1 |
| 18 | *Cacao swollen shoot Togo B virus* | Cacao swollen shoot Togo B virus | CSSToBV | MN179344 |
| 19 | *Cacao yellow vein-banding virus* | Cacao yellow vein-banding virus | CYVBV | NC_033739.1 |
| 20 | *Green Sichuan pepper vein clearing-associated virus* | Green Sichuan pepper vein clearing-associated virus | GSPVCaV | MK371354.1 |
| 21 | *Cacao swollen shoot CD virus* | Cacao swollen shoot CD virus | CSSCDV | NC_038378.1 |
| 22 | *Cacao swollen shoot virus* | Cacao swollen shoot virus | CSSV | NC_001574.1 |

Note, that the first twelve badnaviruses listed in Table 4.2. (1st to 12th) have *simple* replicators exchanging one motif, while replicators from 13th to 22 have c*omplex* replicators exchanging 2 to 4 motifs. All other KM-encoded badnavirus genomes generate replicators with non-trivial sets of transmitted patterns. The four species shown in Table 4.3 have simple single-motif replicators for KM-encoded genomes with maximum size $K=5$, as well as 3-periodic motifs for WS-encoded genomes, and therefore belong to the cell (3,3s). Their hosts are three dicots (Kalanchoe, Rubus, Spiraea) and one monocot (Pineapple).

Table 4.3. Virus genomes in a cell (3, 3s).

| № | Species | Virus name | Abbreviation | Accession |
|---|---|---|---|---|
| 1 | *Kalanchoe top-spotting virus* | Kalanchoe top-spotting virus | KTSV | NC_004540.1 |
| 2 | *Rubus yellow net virus* | Rubus yellow net virus | RYNV | MZ358192 |
| 3 | *Spiraea yellow leafspot virus* | Spiraea yellow leafspot virus | SYLSV | MW080370 |
| 4 | *Pineapple bacilliform CO virus* | Pineapple bacilliform CO virus | PBCOV | LC507821 |

The ten species presented in Table 4.4 contain 3-period replicator patterns for KM-encoded genomes and therefore belong to cell (3,3). Note, that virus genomes belonging to this cell have quite different hosts (7 dicots and 3 monocots) and also contain species with 7 ORF (Dracaena mottle virus). Interestingly, in contrast to the cell rich of wood (3,0), cell (3,3) is rich in berries, containing blackberry, mulberry and grapevine virus genomes. For Blackberry virus F and Mulberry badnavirus 1 RTs are identical, although maximal size simple replicators ($K=15$) have different 3-periodic motifs (Fig. 16).

Table 4.4. Virus genomes in a cell (3, 3).

| № | Species | Virus name | Abbreviation | Accession |
|---|---|---|---|---|
| 1 | *Taro bacilliform virus* | Taro bacilliform virus | TaBF | MG833013 |
| 2 | *Sweet potato papakuy virus* | Sweet potato papakuy virus | SPPV | NC_015655.1 |
| 3 | *Grapevine Roditis leaf discoloration-associated virus* | Grapevine Roditis leaf discoloration-associated virus | GRLDaV | MT783680.1 |
| 4 | *Grapevine badnavirus 1* | Grapevine badnavirus 1 | GBV 1 | NC_055481 |
| 5 | *Blackberry virus F* | Blackberry virus F | BVF | NC_029303.1 |
| 6 | *Mulberry badnavirus 1* | Mulberry badnavirus 1 | MBV1 | NC_026020.2 |
| 7 | *Camellia lemon glow virus* | Camellia lemon glow virus | CLGV | NC_055598.1 |
| 8 | *Dracaena mottle virus* | Dracaena mottle virus | DrMV | NC_008034 |
| 9 | *Cacao bacilliform Sri Lanka virus* | Cacao bacilliform Sri Lanka virus | CBSLV | NC_040809 |
| 10 | *Sugarcane bacilliform Guadeloupe A virus* | Sugarcane bacilliform Guadeloupe A virus | SCBGAV | FJ824813 |

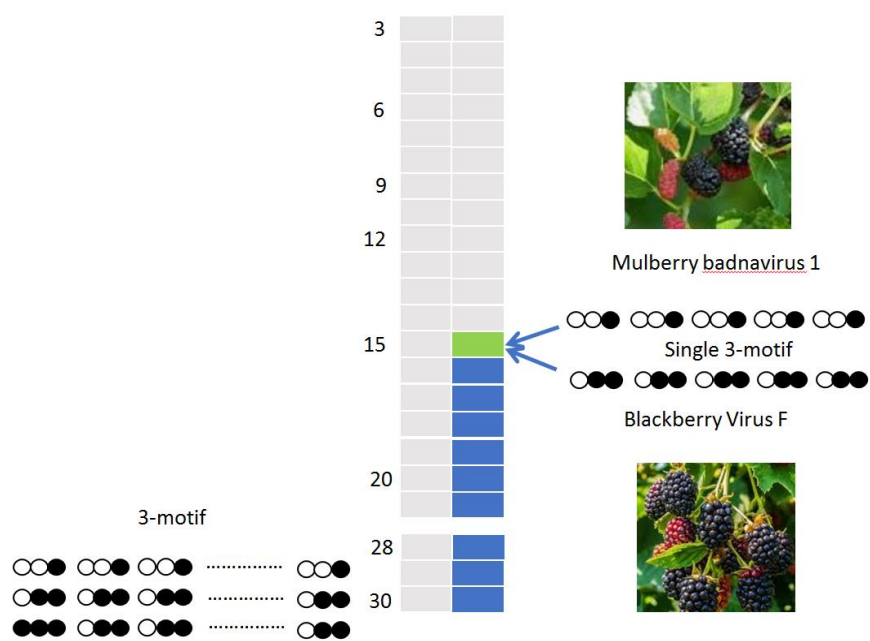

**Fig. 16.** Identical RTs and motifs of badnaviruses of two berries, Mulberry badnavirus 1 (6945 bp) and Blackberry Virus F (7663bp) which belong to the cell (3, 3). Despite the significant difference in sequence lengths, NRA without sequence alignment confidently combines these two badnaviruses into one cell. Moreover, the sets of patterns of WS-encoded genomes for the maximum studied network size ($K$=30) turn out to be identical. The maximum sizes of simple replicators of KM-encoded genomes also coincide ($K$=15), but individual motifs differ from each other.

The eighteen KM-encoded badnavirus genomes shown in Table 4.5 generate replicators with 4-period patterns, and therefore belong to the cell (3,4). Note, that these viruses can cause diseases in different hosts [40].

The first twelve badnaviruses in Table 4.5 (from 1st to 12th) generate simple replicators exchanging a single motif, while viruses from 13th to 18th generate complex replicators. It is noteworthy that, unlike viruses in a cell (3, NoR), only 5 of them have dicotyledonous hosts (cycad leaf necrosis virus is neither dicotyledonous nor monocotyledonous, but rather a gymnosperm plant) - some examples of badnaviruses from this cell are shown in Fig. 17.

Table 4.5. Virus genomes in a cell (3, 4).

| № | Species | Virus name | Abbreviation | Accession |
|---|---|---|---|---|
| 1 | *Badnavirus aucubae* | aucuba ringspot virus | AuRV | LC487411 |
| 2 | *Banana streak GF virus* | banana streak GF virus | BSGFV | NC_007002 |
| 3 | *Banana streak IM virus* | banana streak IM virus | BSIMV | NC_015507 |
| 4 | *Banana streak OL virus* | banana streak OL virus | BSOLV | JQ409540 |
| 5 | *Banana streak UA virus* | banana streak UA virus | BSUAV | NC_015502.1 |
| 6 | *Banana streak UM virus* | banana streak UM virus | BSUMV | NC_015505.1 |
| 7 | *Banana streak VN virus* | banana streak VN virus | BSVNV | KJ013510.1 |
| 8 | *Codonopsis vein-clearing virus* | codonopsis vein-clearing virus | CoVCV | MK044821.1 |
| 9 | *Dioscorea bacilliform SN virus* | Dioscorea bacilliform SN virus | DBSNV | DQ822073.1 |
| 10 | *Dioscorea bacilliform TR virus* | Dioscorea bacilliform TR virus | DBTRV | NC_038995.1 |
| 11 | *Grapevine vein-clearing virus* | grapevine vein-clearing virus | GVCV | NC_015784 |
| 12 | *Cycad leaf necrosis virus* | cycad leaf necrosis virus | CLNV | NC_011097 |
|  |  |  |  |  |
| 13 | *Banana streak UA virus* | banana streak UA virus | BSUAV | NC_015502.1 |
| 14 | *Sugarcane bacilliform Guadeloupe A virus* | sugarcane bacilliform Guadeloupe A virus | SCBGAV | NC_038382.1 |
| 15 | *Sugarcane bacilliform IM virus* | sugarcane bacilliform IM virus | SCBIMV | NC_003031.1 |
| 16 | *Fig badnavirus 1* | fig badnavirus 1 | FBV1 | KT809307.1 |
| 17 | *Dioscorea bacilliform RT virus 3* | Dioscorea bacilliform RT virus 3 | DBRTV3 | MF476845 |
| 18 | *Cacao swollen shoot Ghana Q virus* | cacao swollen shoot Ghana Q virus |  |  |

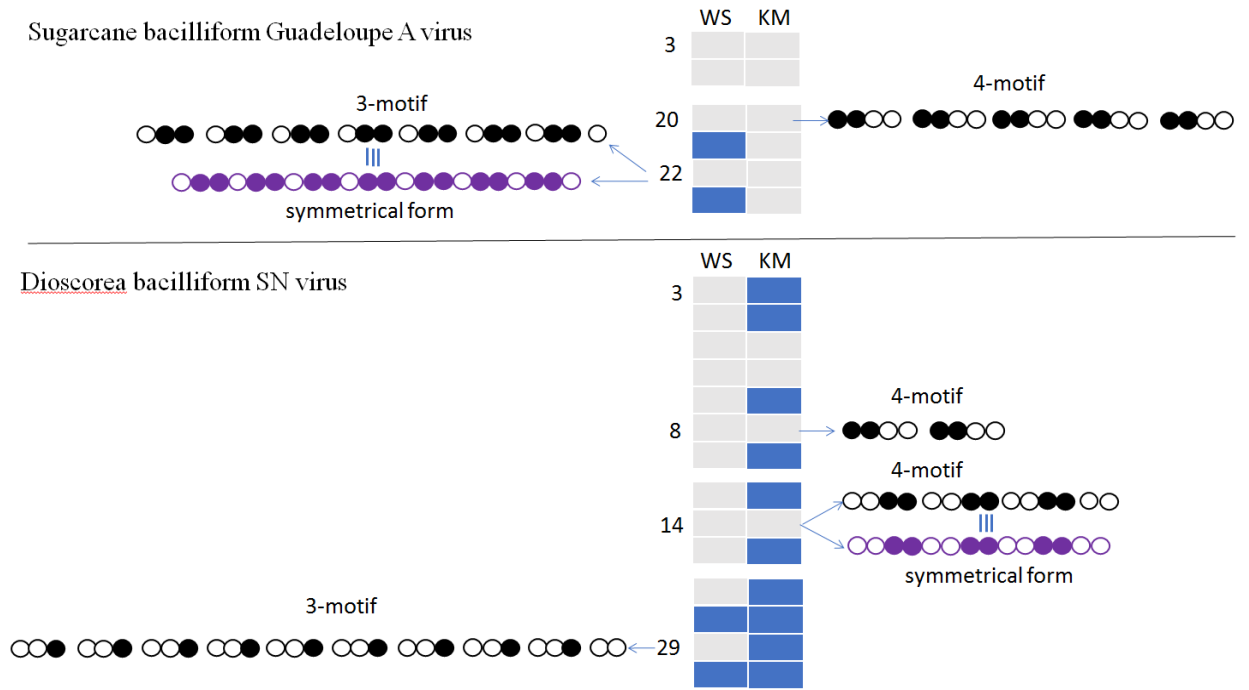

**Fig. 17.** RTs and motifs of Sugarcane bacilliform Guadeloupe A virus and Dioscorea bacilliform SN virus. The presence of 3-motifs for WS-encoded genomes and 4-motifs for KM-encoded genomes is easy to see. However, some periods ($T=3$ for both viruses) and $T=4$ for DBSNV (in the case of network size $K=14$) do not fit into the motif length an integer number of times (for example, the number 14 is not a factor of 4). At the same time, these motifs have a clear symmetry. Next, we will discuss this phenomenon in detail.

This trend also characterizes *intermediate* cell called (3, 3-4) containing viruses with KM-encoded genomes which replicators have 3 or 4-periodic transmitted patterns or patterns with varying parts of 3 and 4 periods (Fig.18). This cell contains 7 viruses (Table 4.6) with only 2 dicot hosts and 5 monocot ones. Only piper yellow mottle virus has a simple single-pattern replicator but, remarkably, three viruses cause yellow mottle disease (the first three in Table 4.6). So, as in the case of a cell (3, NoR) where some viruses of mosaic disease are present, NRA shows the potential ability to place some other virus diseases in one cell (here (3, 3-4)).

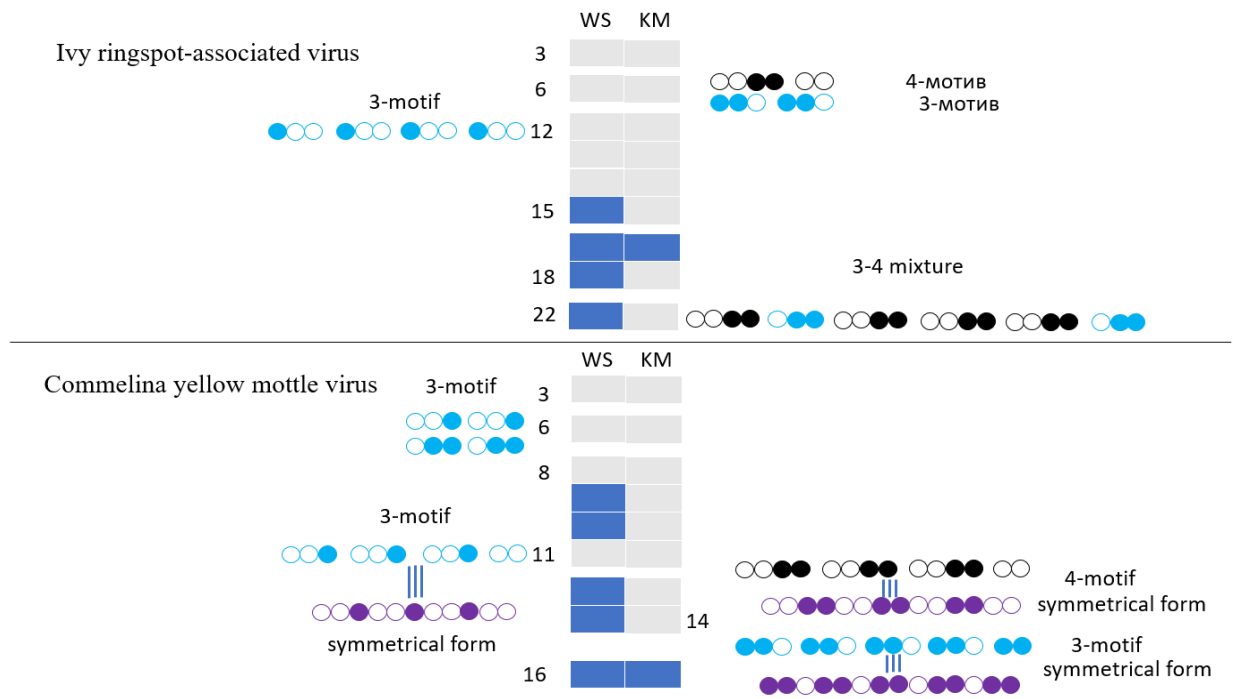

**Fig. 18.** RTs and motifs of Ivy ringspot-associated virus (top) and Commelina yellow mottle virus (bottom). The former has a mixed replicator patterns consisting of 3-period (stained in white and blue) and 4-period (stained in white and black) regions for the KM-encoded genome, while the latter contains complex replicators having both 3-periodic and 4-periodic patterns. The symmetrical forms of such patterns are shown in white and purple.

Table 4.6. Virus genomes in cell (3, 3-4).

| № | Species | Virus name | Abbreviation | Accession |
|---|---|---|---|---|
| 1 | *Piper yellow mottle virus* | Piper yellow mottle virus | PYMoV | MW116776 |
| 2 | *Canna yellow mottle virus* | Canna yellow mottle virus | CaYMV | MF074075.1 |
| 3 | *Commelina yellow mottle virus* | Commelina yellow mottle virus | ComYMV | NC_001343.1 |
| 4 | *Dioscorea bacilliform RT virus 1* | Dioscorea bacilliform RT virus 1 | DBRTV1 | NC_038986.1 |
| 5 | *Dioscorea bacilliform RT virus 2* | Dioscorea bacilliform RT virus 2 | DBRTV2 | NC_038987.1 |
| 6 | *Banana streak MY virus* | banana streak MY virus | BSMYV | KR014107 |
| 7 | *Ivy ringspot-associated virus* | Ivy ringspot-associated virus | IRSaV | NC_055604 |

Consider other cells for badnavirus genomes. Two KM-encoded genomes of two viruses of monocot Dioscorea (Dioscorea bacilliform AL virus, Dioscorea bacilliform ES virus) generate replicators with periods 6,7 and 9 (Tables 4.7 and 4.8). So, for these cells (3, 6-7) and (3,9) there are no badnaviruses with dicot hosts.

Table 4.7. Virus genomes in cell (3, 6-7).

| № | Species | Virus name | Abbreviation | Accession |
|---|---|---|---|---|
| 1 | *Dioscorea bacilliform AL virus* | dioscorea bacilliform AL virus | DBALV | NC_038381.1 |

Table 4.8. Virus genomes in cell (3, 9).

| № | Species | Virus name | Abbreviation | Accession |
|---|---|---|---|---|
| 1 | *Dioscorea bacilliform ES virus* | dioscorea bacilliform ES virus | DBESV | KY827394 |

Finally, three badnaviruses − polyscias mosaic virus (PoMV, MH475918), sugarcane bacilliform Mo virus (SCBMOV, M89923) and gooseberry vein banding associated virus (GVBaV, JQ316114) – do not have distinct periods for KM-encoded genomes that generate complex replicators with patterns that have approximately monotonously decaying neuron activity. Possibly, this means the random character of bits in KM-encoded virus genomes [23] and may be related to the transition to the spin glass phase in the Hopfield neural network [6, 41]. This situation shows that in addition to describing replicators in terms of *periodicity* it is advisable to introduce such characteristic of replicator pattern sets as *monotony* (M). Thus, the cell to which such viruses belong can be called (3, M).

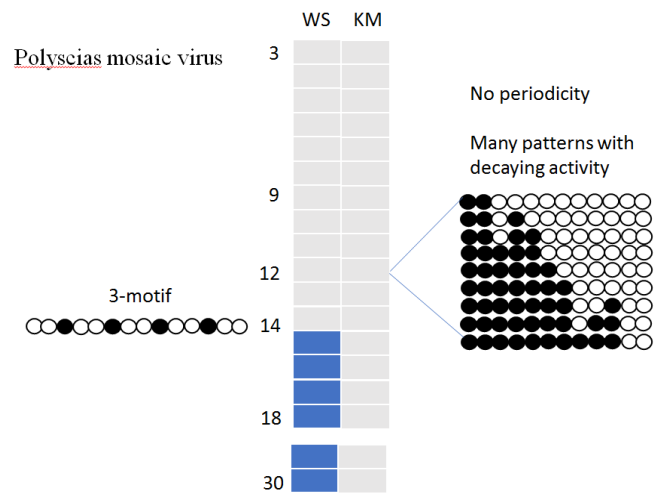

**Fig. 19.** RTs and motifs of Polyscias mosaic virus. The replicators generated using the KM-encoded genome have a complex set of transmitted patterns (9 for *K*=12) with monotonously decaying activity (decrease in the number of active neurons – white circles).

Taking into account these cases and, considering that in some (relatively rare) cases, the periodicity of motifs may be absent, we can show the cells occupied by badnaviruses in Fig. 20.

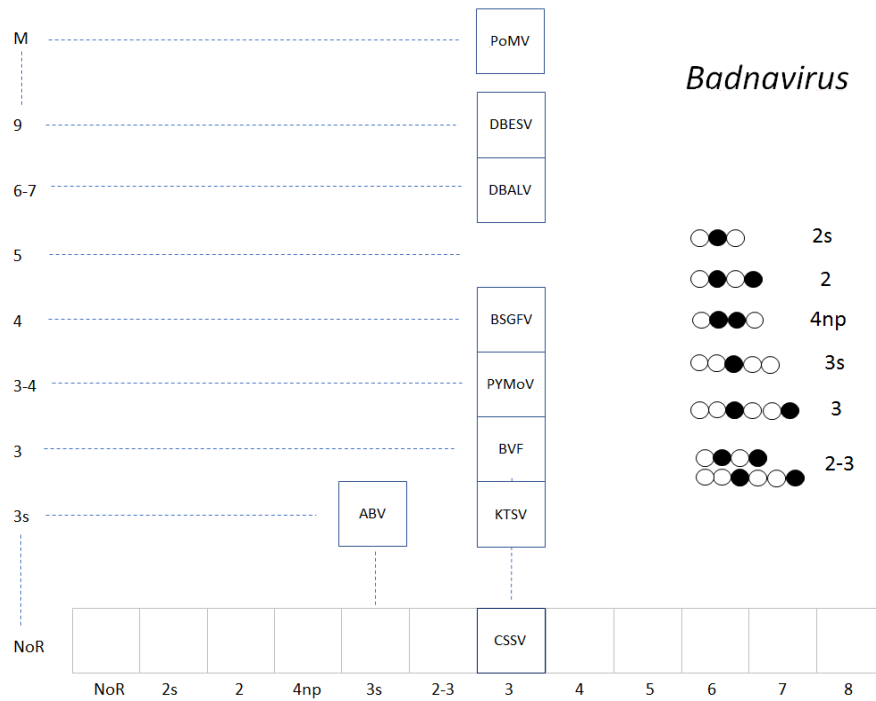

**Fig. 20.** Binomial table cells occupied by badnavirus genomes. Each cell shows one typical virus. With the exception of ABV only, all members of the genus *Badnavirus* are placed in column (3, *), where replicators with 3-periodic patterns for WS-encoded viral genomes are located.

The analysis of this table reveals interesting qualitative dependences of the characteristics of the virus host and the disease on the position in the column (3, *) of the table − Fig.21. First, an increase in the period of replicators obtained using KM-encoded genome from 3 to 9 corresponds to the subsequent average change of hosts from trees through berries and liana to grass and root crops. Also, the portion of dicots among the hosts decreases correspondingly from more than 80% to 0%. Viruses from the same host, especially cacao and banana, are presumably concentrated in the same cells ((3, NoR) and (3,4), correspondingly). This trend continues for types of viral disease (for example, yellow mottle disease is caused by viruses which genomes are concentrated in the cell (3, 3-4)). It can be concluded, that some properties associated with phenotypic characteristics of viruses are associated with the position of their genomes in the presented binomial table.

| | Plant type | Typical host | Dicot | Monocot | Dicot, % | Typical disease |
|---|---|---|---|---|---|---|
| 6-7, 9 | Root vegetable | Yam | 2 | 0 | 0% | |
| 4 | Grass | Banana | 5 | 12 | 28% | Vein-clearing |
| 3-4 | Liana | Ivy | 2 | 5 | 29% | Yellow mottle |
| 3 | Berry | Blackberry | 10 | 4 | 67% | |
| NoR 3 | Tree | Cacao | 18 | 4 | 82% | Mosaic |

**Fig. 21.** Some phenotypic characteristics of members of the *Badnavirus* genus are partly related to the position of the viral genome in the binomial table.

Let us make some remarks about the complexity of the replicator and the possible further refinement of the classification. It is clear that a simplified classification scheme for the virus genome based on the periodicity of replicator motifs for two coding schemes cannot be sufficient in the general case. Although it can be applied to most viral genomes, there are many interesting complex cases that should be carefully considered.

*Some remarks on periodic and symmetrical-periodic motifs*

When we find clear evidence for the existence of periods in the replicator motif, we can nonetheless see variations on this situation. If $K$ is the length of the motif (net size) and $T$ is the period found in the motif, then it is possible that this motif contains an integer number $n$ of periods $K=nT$. For example, for $K=9$ and $T=3$ this motif may look like (11-1  11-1  11-1).   We will call such motifs periodic. On the other hand, a motif (a pattern transmitted by a replicator) may contain a non-integer number of periods, which nevertheless are clearly distinguishable in the whole motif, sometimes having a symmetrical shape. So, for $K=8$ and $T=3$ it can look like    (11 – 1 11 -1 11). We will call such motifs symmetrically periodic motifs. The reason for distinguishing between these two cases is that in the entire set of genomic motifs, we may encounter interesting interactions between them. Let's look at some specific examples. In the case of human papillomaviruses, the motifs of the  $α$-2 and $α$-4 species have periods $T=2$, as well as an even and odd number of components, $K$, which gives a strictly periodic and symmetrical forms.

For example, type HPV2 has motifs (1 -1 1 -1), (1 -1 1 -1 1), (1 -1 1 -1 1 -1), (1 -1 1 -1 1 -1 1), (1 -1 1 -1 1 -1 1 -1), (1 -1 1 -1 1 -1 1 -1 1). Another example of a virus belonging to the *Caulimoviridae* family: the set of genome motifs of taro bacilliform virus isolate Tz24 (MG833013) contains 9 motifs of the WS-encoded genome (from 3 to 11). The motif of length K=9 is periodic with period T=3: (1 1 -1 1 1 -1 1 1 -1), and the motif of length K=11 is symmetrically-periodic: (1 1 - 1 1 1 -1 1 1 -1 1 1 ).

But there are also very intriguing cases when periodic and symmetrically-periodic motifs seem *incompatible*. For example, the RT of crow polyomavirus CPyV (DQ192570) contains only symmetrically-periodic WS-encoded motifs with period $T=3$ and lacks $K=6, 9, 12, 15, 18$ size replicators that could potentially have 3-periodic motifs with integer number of periods. Also, another avian polyomavirus Adelie penguin polyomavirus - AdPyV_Crozier (KP033140.2) – has a similar RT without replicators of sizes 9, 12, 15, 18 (Fig. 22). It is remarkable that a very similar picture is observed in the case of a plant virus belonging to the family *Geminiviridae* (chapter 5). The DNA-A of the bipartite Allamanda leaf mottle distortion virus AllLMoDV (KC202818) is characterized by RT, which also shows the absence of replicators having sizes 6, 9, 12, 15, 18. So only symmetrical-periodic motifs survive. Note, that the genomes of these two avian and plant viruses are very different in lengths, 5079 bp and 2772 bp, respectively. Thus, it is quite difficult to align them to establish sequence similarity. On the other hand, NRA clearly demonstrates their *qualitative* similarity. What meaning can be seen in this? One possible interpretation can be imagined. Allamanda is one of the plants that are dangerous for birds, as it contains poisons that lead to weakening, illness and death of the bird. Crows have also been known to exhibit cannibalism towards other weak and alien crows. Thus, some routes for transmission of the virus may take place.

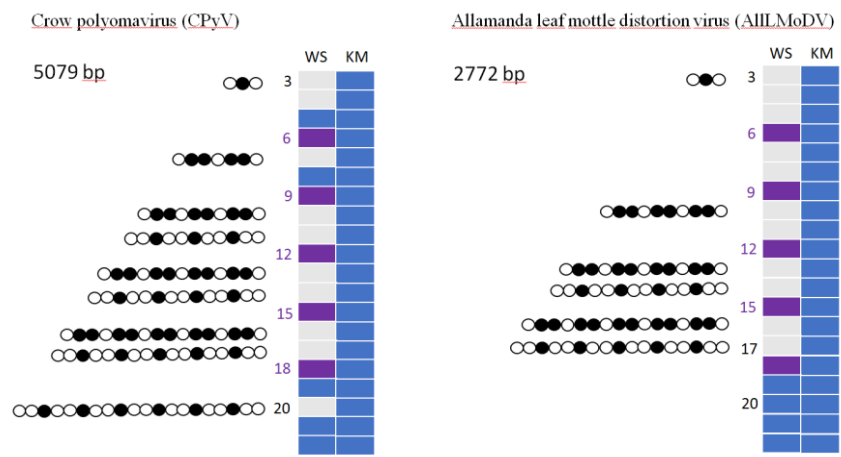

**Fig. 22.** RTs of crow polyomavirus CPyV (left) and Allamanda leaf mottle distortion virus AllLMoDV (right) are very similar and contain only symmetrically-periodic WS-encoded motifs with period $T=3$ and demonstrate the absence of replicators of size $K=6, 9, 12, 15, 18$ (shown in purple) that could potentially have 3-period motifs with an integer number of periods.

Remind, that members of the genus *Badnavirus* have the form of a bacillus. Apart from the genus *Tungrovirus* viruses of all other genera of the family *Caulimoviridae* have an isometric form.

Genus: *Caulimovirus*

This genus contains 11 members, three of them belong to cell (NoR, NoR).

Table 4.9. Virus genomes in cell (NoR, NoR).

| № | Species | Virus name | Abbreviation | Accession |
|---|---|---|---|---|
| 1 | *Angelica bushy stunt virus* | Angelica bushy stunt virus | AnBSV | NC_043523.1 |
| 2 | *Carnation etched ring virus* | Carnation etched ring virus | CERV | NC_003498.1 |
| 3 | *Figwort mosaic virus* | Figwort mosaic virus | FMV | NC_003554 |

Dahlia mosaic virus can be certainly put to cell (3s, NoR).

Table 4.10. Viruses genomes in cell (3s, NoR).

| № | Species | Virus name | Abbreviation | Accession |
|---|---|---|---|---|
| 1 | *Dahlia mosaic virus* | Dahlia mosaic virus | DMV | NC_018616 |

The Mirabilis mosaic virus has, in addition to a single-pattern replicator typical for a cell (3s, NoR) (1 1 -1 1 1), a *very specific pattern* for $K=11$ with a single central "off" state of the neuron (Fig. 23). Such a special symmetrical state can be called 6s (6=5+1, $K=2\cdot5+1$). Note that this will be in direct correspondence with the notation 2s (2=1+1, $K=2\cdot1+1$) and 3s (3=2+1, $K=2\cdot2+1$), because both of the latter cases also include motifs with a single "off" state of the neuron. Therefore, we place the Mirabilis mosaic virus in a new cell called (6s, NoR).

Table 4.11. Virus genomes in cell (6s, NoR).

| № | Species | Virus name | Abbreviation | Accession |
|---|---|---|---|---|
| 2 | *Mirabilis mosaic virus* | Mirabilis mosaic virus | MMV | NC_004036 |

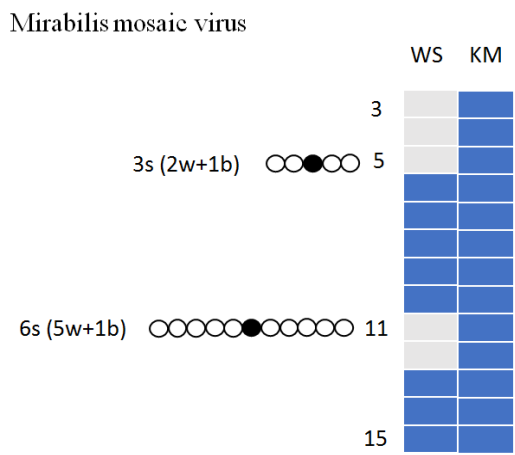

**Fig. 23.** RT of the Mirabilis mosaic virus has a special symmetrical motif K=11. To unite its specification to the familiar 3s motif the notation 6s is introduced. Here 6 is half the sum of the components +1 plus one component corresponding to -1.

Also soybean mild mottle pararetrovirus (NC_018505.1) belongs to cell (3, 4np).

Table 4.12. Virus genomes in cell (3, 4np).

| № | Species | Virus name | Abbreviation | Accession |
|---|---|---|---|---|
| 1 | *Soybean mild mottle pararetrovirus* | soybean mild mottle pararetrovirus | SPuV | NC_018505.1 |

The four other caulimoviruses listed in Table 4.13 belong to the cell (3, NoR).

Table 4.13. Virus genomes in cell (3, NoR).

| № | Species | Virus name | Abbreviation | Accession |
|---|---|---|---|---|
| 1 | *Atractylodes mild mottle virus* | atractylodes mild mottle virus | AMMV | MZ029050 |
| 2 | *Horseradish latent virus* | horseradish latent virus | HRLV | NC_018858 |
| 3 | *Lamium leaf distortion associated virus* | lamium leaf distortion associated virus | LLDAV | NC_010737.1 |
| 4 | *Strawberry vein banding virus* | strawberry vein banding virus | SVBV | MT731326 |

Finally, the famous cauliflower mosaic virus, the first plant virus to contain DNA rather than RNA as its genetic material, must be placed in a cell (3,3).

Table 4.14. Virus genomes in cell (3, 3).

| № | Species | Virus name | Abbreviation | Accession |
|---|---|---|---|---|
| 1 | *Cauliflower mosaic virus* | cauliflower mosaic virus | CaMV | V00140 |

Genus: *Cavemovirus*

In this genus, all viruses have in common that they do not have replicators for their WS-encoded genomes. The epiphyllum virus 4 isolate Ebert also lacks replicators for the KM-encoded genome and thus occupies a cell (NoR, NoR).

Table 4.15. Virus genomes in cell (NoR, NoR).

| № | Species | Virus name | Abbreviation | Accession |
|---|---|---|---|---|
| 1 | *Epiphyllum virus 4 isolate Ebert* | epiphyllum virus 4 isolate Ebert | EpV-4 | NC_055588.1 |

Cassava vein mosaic virus is located in the nearest cell (NoR, 2s).

Table 4.16. Virus genomes in cell (NoR, 2s).

| № | Species | Virus name | Abbreviation | Accession |
|---|---|---|---|---|
| 1 | *Cassava vein mosaic virus* | Cassava vein mosaic virus | CsVMV | NC_001648.1 |

A sweet potato caulimo-like virus occupies an *intriguing* new cell (NoR, 2) having a short ($K=4$) 2-period motif for the KM-encoded genome.

Table 4.17. Virus genomes in cell (NoR, 2).

| № | Species | Virus name | Abbreviation | Accession |
|---|---|---|---|---|
| 1 | *Sweet potato caulimo-like virus* | Sweet potato caulimo-like virus | SPCV | NC_015328.1 |

Genus: *Soymovirus*

In this genus we find a second virus belonging to the cell (NoR, 2) – the peanut chlorotic streak virus, for which 2-periodicity characterizes the replicators of KM-encoded genome. Unlike the short replicator of SPCV, the maximum length of such replicators is $K=10$.

Table 4.18. Virus genomes in cell (NoR, 2).

| № | Species | Virus name | Abbreviation | Accession |
|---|---|---|---|---|
| 1 | *Peanut chlorotic streak virus* | peanut chlorotic streak virus | PCSV | NC_001634.1 |

Also, Blueberry red ringspot virus enters a still empty cell (NoR, 3s).

Table 4.19. Virus genomes in cell (NoR, 3s).

| № | Species | Virus name | Abbreviation | Accession |
|---|---|---|---|---|
| 1 | *Blueberry red ringspot virus* | blueberry red ringspot virus | PCSV | NC_003138.2 |

The genomes of two other viruses of this genus are located in densely populated cells (NoR, NoR) and (3s, NoR).

Table 4.20. Virus genomes in cell (NoR,NoR).

| № | Species | Virus name | Abbreviation | Accession |
|---|---|---|---|---|
| 1 | *Soybean chlorotic mottle virus* | soybean chlorotic mottle virus | SbCMV | NC_001739 |

Table 4.21. Virus genomes in cell (3s, NoR).

| № | Species | Virus name | Abbreviation | Accession |
|---|---|---|---|---|
| 1 | *Cestrum yellow leaf curling virus* | cestrum yellow leaf curling virus | CmYLCV | NC_004324.3 |

Genus: *Tungrovirus*

Like all members of the genus *Badnavirus,* the only member of the genus *Tungrovirus,* the Rice tungro bacilliform virus, also has the form of a bacillus. Members of the genus *Badnavirus* can be distinguished from it by genome organization, the lack of any RNA splicing during replication, etc. NRA reveals *almost 2- periodic* motif of the length $K=10$ (as for exactly 2-periodic motif of SPCV and PCSV) for KM-encoded genome, and we can *conditionally* put it in cell (NoR, 2) (Fig. 24). Note, that despite of the same bacillus form, this cell is not occupied by any member of the genus *Badnavirus*. So, NRA also reliably separates two genera – *Badnavirus* and *Tungrovirus*.

Table 4.22. Virus genomes in cell (NoR, 2).

| № | Species | Virus name | Abbreviation | Accession |
|---|---|---|---|---|
| 1 | *Rice tungro bacilliform virus* | Rice tungro bacilliform virus | RTBV | AF220561 |

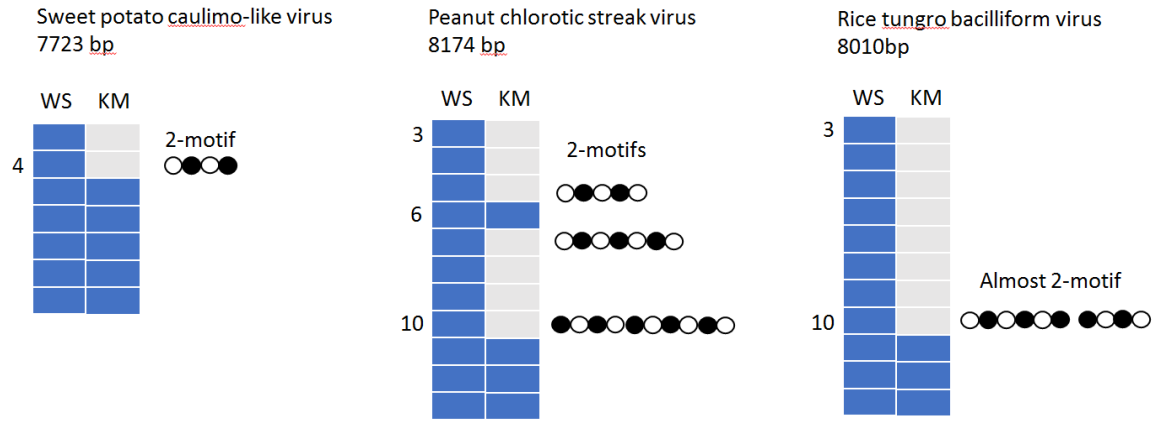

**Fig. 24.** RTs of Sweet potato caulimo-like virus (genus *Cavemovirus*), Peanut chlorotic streak virus (genus *Soymovirus*) and Rice tungro bacilliform virus (genus *Tungrovirus*). Note that they are mirror images of RTs of papillomaviruses that cause warts.

Genus: *Petuvirus*

The petunia vein clearing virus is housed in a cell rich in badnaviruses

Table 4.23. Virus genomes in cell (3, NoR).

| № | Species | Virus name | Abbreviation | Accession |
|---|---|---|---|---|
| 1 | *Petunia vein clearing virus* | petunia vein clearing virus | PVCV | NC_001839.2 |

Genus: *Rosadnavirus*

The rose yellow vein virus has a RT very similar to that of the mirabilis mosaic virus and completely identical replicators (Fig.25). So, apparently, it can be placed in the cell (6s, NoR).

Table 4.24. Virus genomes in cell (6s, NoR).

| № | Species | Virus name | Abbreviation | Accession |
|---|---|---|---|---|
| 1 | *Rose yellow vein virus* | rose yellow vein virus | RYVV | NC_020099 |

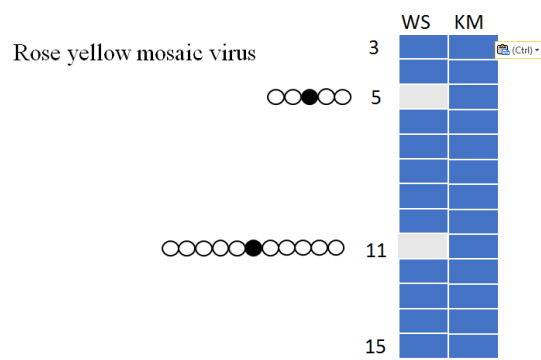

**Fig. 25.** The RT of the rose yellow mosaic virus is similar to that of mirabilis mosaic virus shown in Fig.23.

Genus: *Solendovirus*

Two viruses of this genus have identical RTs, suggesting that the NRA may reflect the similarity of viruses belonging to a genus defined using standard approaches. It is also notable that these two viruses cause the same *vein clearing disease*, and this is also indicative of the NRA's ability to place common disease viruses in the same cell, here in a cell (NoR, 2s).

Table 4.25. Virus genomes in cell (NoR, 2s).

| № | Species | Virus name | Abbreviation | Accession |
|---|---|---|---|---|
| 1 | *Tobacco vein clearing virus* | tobacco vein clearing virus | TVCV | NC_003378.1 |
| 2 | *Sweet potato vein clearing virus* | sweet potato vein clearing virus | SPVCV | NC_015228 |

Genus: *Dioscovirus*

The only member of this genus does not have replicators for both for WS- and KM-encoded genomes and therefore occupies a cell (NoR, NoR).

Table 4.26. Virus genomes in cell (NoR, NoR).

| № | Species | Virus name | Abbreviation | Accession |
|---|---|---|---|---|
| 1 | *Dioscorea nummularia-associated virus* | dioscorea nummularia-associated virus | DNUaV | MG944237 |

Genus: *Vaccinivirus*

The blueberry fruit drop associated virus is the only member of this genus. Its genome occupies a cell (3, NoR) typical for badnaviruses.

Table 4.27. Virus genomes in cell (3, NoR).

| № | Species | Virus name | Abbreviation | Accession |
|---|---|---|---|---|
| 1 | *Blueberry fruit drop associated virus* | Blueberry fruit drop associated virus | BFDaV | KT148886.1 |

Genus: *Ruflodivirus*

Finally, rudbeckia flower distortion virus is also the only member of this genus. Its genome occupies a cell (3s, NoR) common to many papillomaviruses, polyomaviruses and caulimoviruses.

Table 4.28. Virus genomes in cell (3s, NoR).

| № | Species | Virus name | Abbreviation | Accession |
|---|---|---|---|---|
| 1 | *Rudbeckia flower distortion virus* | rudbeckia flower distortion virus | RuFDV | NC_011920.1 |

To summarize, cells occupied by genomes of viruses of the *Caulimoviridae* family are schematically shown in Fig.26

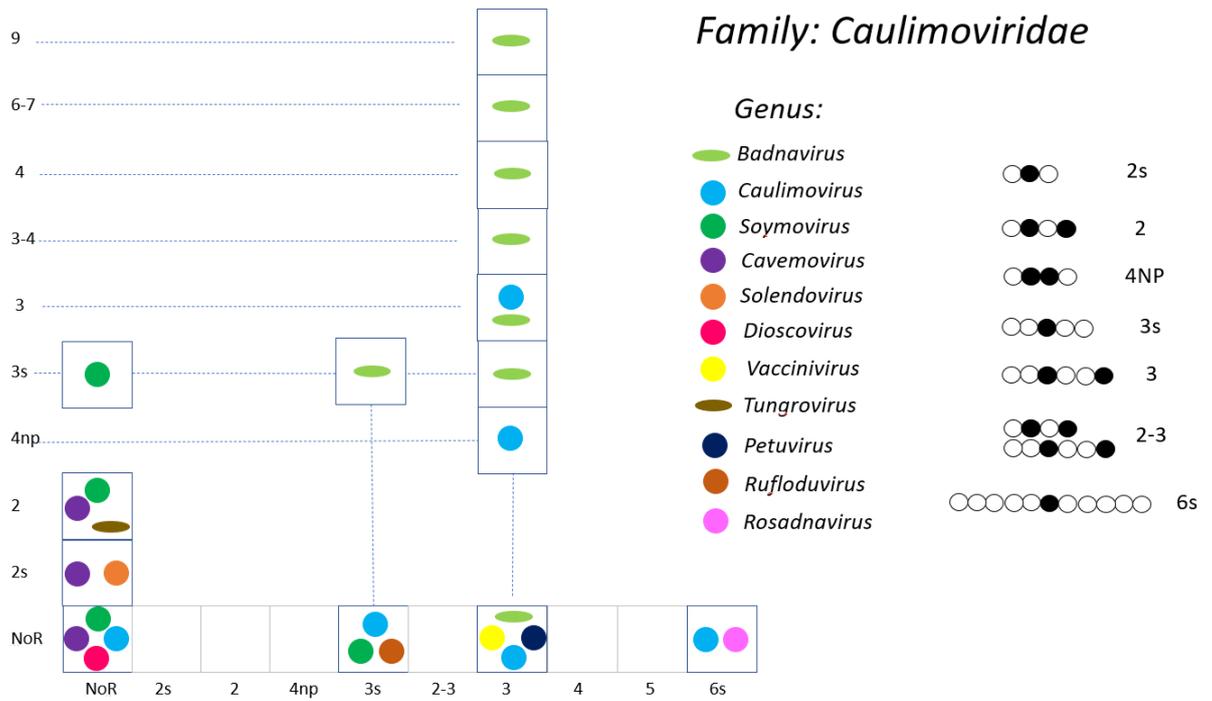

**Fig. 26.** Cells of the binomial table that are occupied by the genomes of viruses belonging to the family *Caulimoviridae.*

**5. Some examples of application of NRA to the members of *Geminiviridae* family**

Here we present some examples of virus genomes from the family *Geminiviridae,* which can be attributed to so far empty cells not occupied by members of *Papillomaviridae, Polyomaviridae* and *Caulimoviridae*.

Note, that some members of the family *Caulimoviridae* do not have replicators for WS-encoded genomes, but they do for KM-encoded ones, so they fill in the cells of the first column (NoR, *) of the virus genome table. The remaining new cells from this column are occupied by the genomes of the family *Geminiviridae*. It is noteworthy that viruses of its genus *Begomovirus* can have monopartite genome having one DNA and well as bipartite genome with two DNAs (DNA-A and DNA-B). Of course, this will be a problem if we want to use the NRA to classify viruses. But since we are trying to build a binomial classification of viral genomes this problem disappears.

Bipartite abutilon mosaic Bolivia virus resides in a column (NoR,*) having 5- and 10-period motifs for the KM-encoded DNA-A while its DNA-B generates a replicator with 5-period motif. In addition, monopartite abutilon golden mosaic Yucatan virus has a 6-period replicator for the KM-encoded genome and 3-period replicator for the WS-encoded genome.

In presenting remarks on the complexity of the replicator and refinement of the classification, we have already reviewed the case of allamanda leaf mottle-distortion virus belonging to the genus *Begomovirus* and cell (3, NoR). Here, we note that the DNA-B of the genome of this virus generates a replicator directing this genome into the cell (NoR, M).

The DNA-B of the asystasia mosaic Madagascar virus generates replicators that direct it in a new cell (3s,3). DNA-A of tomato mosaic Havana virus is placed in a new cell (3,2) and DNA-B of tomato mosaic Havana virus is placed in the new cell (3,M). DNA-A of corchorus yellow spot virus is also placed in a cell (3,2), while DNA-B of this virus – in the cell (3,10)! Note, however, that many other members of the genus *Begomovirus* have a complex mixture of motifs of different length from 5 to 9 and can hardly be placed in any cell of the constructed table. For example, DNA-B of clerodendron golden mosaic virus should be simultaneously placed into (NoR, 6) and (Nor, M) cells (Fig. 26) while DNA-B of Wissadula golden mosaic St Thomas both in cells (3,8) and (3,9), etc. Of course, the last virus can be placed in a new intermediate cell (3, 8-9) – we have already introduced similar cells such as (2-3, NoR) and (3, 3-4) for other viruses. But we can also use for such cases the representation adopted in Fig.27, The optimal choice of virus placement in the binomial table can be made after additional research. Note, however, that there is also a radical pluralistic point of view, according to which some viruses can belong to more than one species, while others don't form species at all and "*should be classified using new reticulate categories*" [13].

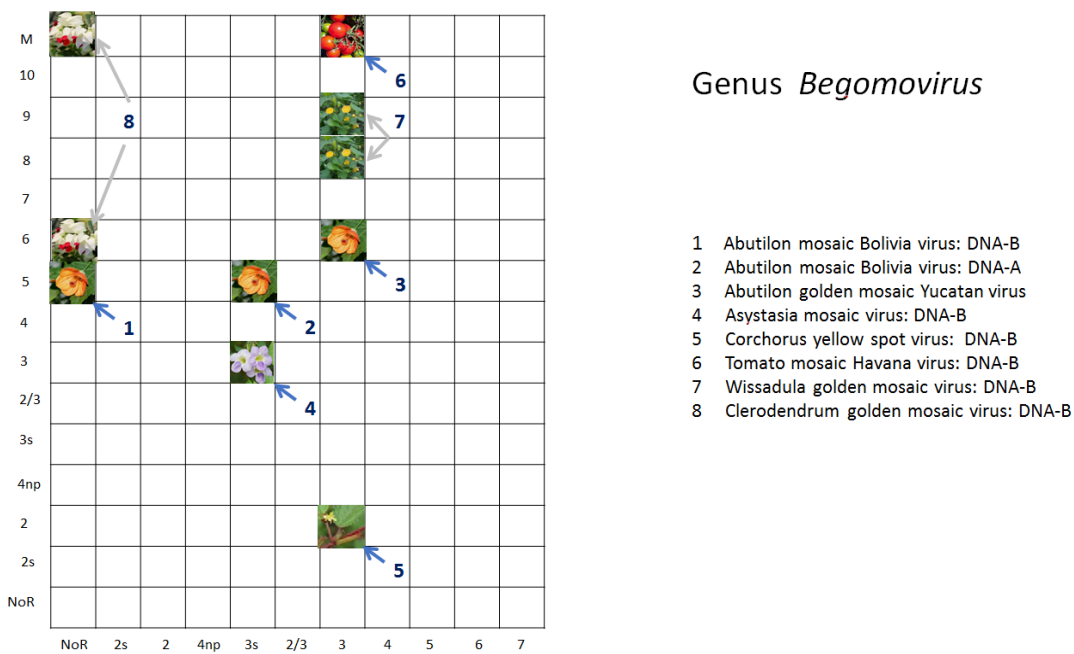

**Fig. 27.** Cells of the binomial table filled by the DNA-A and DNA-B of the viruses belonging to the genus *Begomovirus* of the family *Gemiviridae*. KM-encoded DNA-B of Wissadula golden mosaic virus (7) and DNA-B of Clerodendron golden mosaic virus (8) generate complicated replicators that link them to two different cells of the table.

Table 5.1. The virus genomes occupying the cells shown in Fig. 27.

| № | Species | Virus name | Abbreviation | Accession |
|---|---------|------------|--------------|-----------|
| 1 | *Abutilon mosaic Boliviavirus* | abutilon mosaic Bolivia | AbMBoV | DNA-A: HM585445 DNA-B: HM585446 |
| 2 | *Abutilon golden mosaic Yucatan* | abutilon golden mosaic Yucatan virus | AbGMV | DNA-A: KC430935 |
| 3 | *Allamanda leaf mottle distortion virus* | allamanda leaf mottle distortion virus | AllLMoDV | DNA-A: KC202818 DNA-B: MG969497 |
|   | *Asystasia mosaic Madagascar virus* | asystasia mosaic Madagascar virus | AsMMV | DNA-A: KP663485 DNA-B: KP663484 |
|   | *Tomato mosaic Havana virus* | tomato mosaic Havana virus | ToMHaV | DNA-A: Y14874 DNA-B: Y14875 |
|   | *Corchorus yellow spot virus* | corchorus yellow spot virus | CoYSV | DNA-A: DQ875868 DNA-B: DQ875869 |
|   | *Clerodendron golden mosaic virus* | clerodendron golden mosaic virus | ClGMV | DNA-A: DQ641692 DNA-B: DQ641693 |
|   | *Wissadula golden mosaic virus* | wissadula golden mosaic St Thomas | WGMV | DNA-A: DQ395343 DNA-B: EU158095 |

## 6. Joint Virus Genome Table (VGT)

Now we can combine the data obtained from the study of viruses of the *Papillomaviridae*, *Polyomaviridae*, *Caulimoviridae* and partly *Geminiviridae* families into a common virus genome table (VGT). Of course, it will only contain a small fraction of known viruses. Note that we only studied circular DNA viruses, but NRA can be used for other viruses, including ssRNA, dsRNA, and genes of the viral genome, and this table will have more new filled cells (to illustrate this, we added to the table corresponding data for 4 members of the Mitoviridae family, including 3 complete genomes and one RdRp (see Fig. 28 for details). We also note that the task of placing the virus genome in a given cell is in some cases rather complicated, and this table should generally have a fuzzy form. However, for some important families of viruses that cause disease in humans and animals, this can be done easily, so some observations can be made based on the position of the virus genome in the VGT. The first, but perhaps insufficiently substantiated conclusion at this stage, is that the genomes of animal viruses are located mainly in the cells belonging to the bottom line of the table (*, NoR). On the other hand, plant viruses usually have replicators for their KM-encoded genomes and are therefore distributed over the top rows of cells. Other interesting observations related to the "meeting" of animal and plant virus genomes in one cell (3, NoR) were discussed earlier.

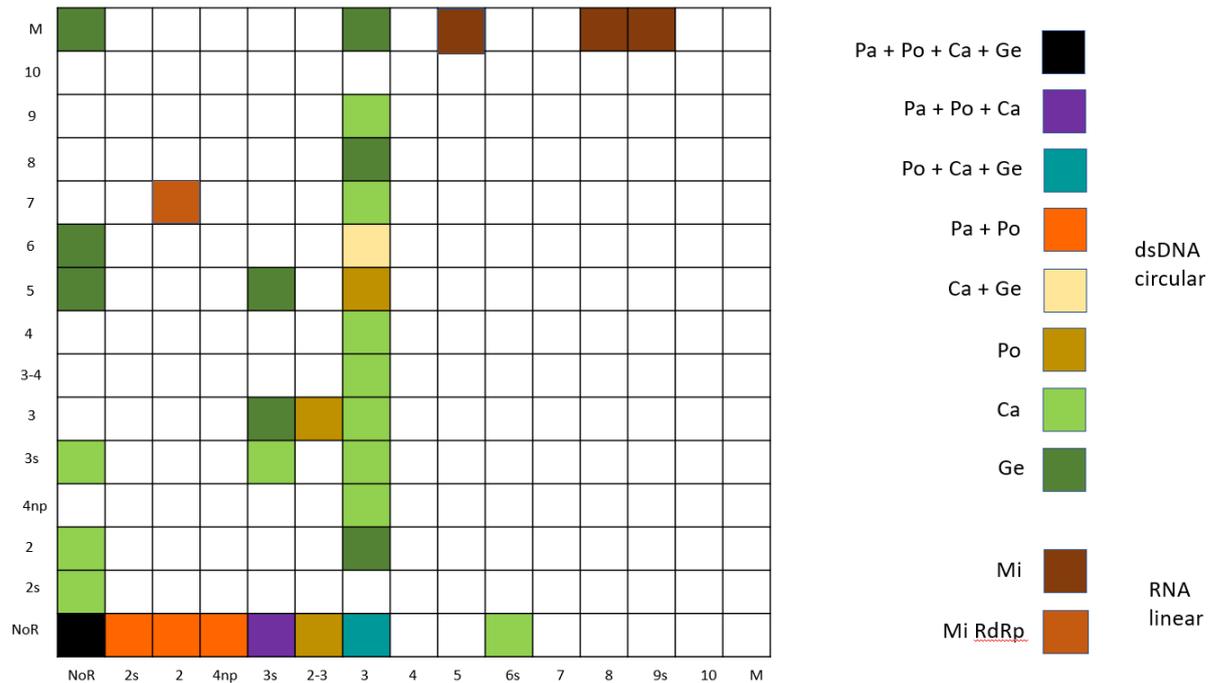

**Fig. 27.** Joint virus genome table showing cells populated with NRA data obtained for families *Papillomaviridae* (human) – Pa, *Polyomaviridae* – Po, *Caulimoviridae* – (Ca), *Geminiviridae* – Ge and also *Mitoviridae* – Mi (Gigaspora margarita mitovirus 1, NC_040702.1 – (5, M), Cronartium ribicola mitovirus 5, NC_030399.1 –(8, M); Fusarium poae mitovirus 4, NC_030864.1 – (9s, M) ; Rhizoctonia mitovirus 1 RdRp, NC_040563.1 – (2, 7) ). Some cells are filled with genomes of viruses belonging to different families,

**Discussion**

We conducted this study to analyze the potential usefulness of the NRA proposed in [19] for constructing a binomial classification of virus genomes based only on knowledge of their complete genomic sequences, without involving other data on phenotype, functions, encoded proteins, etc., and also without the need to align genomic sequences. Comparison of genomic sequences plays an important role in the taxonomy of viruses to distinguish between types, species, genera and families, so we hope that NRA applied to virus genomes can, in principle, provide some additional information for virus classification. We have demonstrated that the periodicity of replicator patterns can be used to organize most viral genomes into a rectangular table to obtain their binomial classification. The key factor is the possibility of independent analysis of two binary sequences representing WS-encoded and KM-encoded genomes. This approach makes it possible to combine the genomes of viruses that are far from each other in terms of the similarity of aligned nucleotide sequences (as in the case of representatives of the *α*- and *v*-human papillomaviruses, and even in the case of viruses belonging to different kingdoms, such as the crow polyomavirus and allamanda leaf mottle distortion virus) or, on the other hand, separate them when they have similar aligned genomes), as in the case of *v* human papillomavirus and porcupine papillomavirus *σ*.

We have also demonstrated that NRA may, to some extent, reflect such general characteristics of the viral phenotype as:

- virus hosts and their possible interconnections (trees in a cell (3, NoR) for the genus *Badnavirus*, rodents in a cells (2-3, NoR) and (3, NoR) and birds in a cell (3,5) for the family *Polyomaviridae*);
- the form of disease (e.g., mosaic disease – cell (3, NoR) or yellow mottle disease –cell (3, 3-4) caused by members of genus *Badnavirus*);
- oncogenicity of the virus (tendency to the absence of neural replicators for both genome coding schemes);
- the form of the epithelium affected by the virus (skin or mucous) (for the *human papillomavirus*);
- morphology (members of the genus *Badnavirus* with bacillary geometry are presumably located in the same column of the VGT - (3,*)).

The principal question for the NRA approach is "*Why does it work?*" requires future research. The general understanding is that neural networks allow for non-linear data transformation, which proves to be very useful in many applications, including data categorization, classification, and pattern recognition. In addition, being complex systems, they have emergent properties and exhibit emergent patterns: for example, these are fine-grained patterns transmitted by replicators that have special shapes, including periodic ones. In any case, the world of known viral genomes is so large that there remains a wide field for future research on the application of NRA and the "artificial pathogen" (neural replicator) model [19] of genomic sequences and to evaluate their usefulness. Perhaps, NRA may be particularly useful for the analysis of metagenomic data. Actually NRA permits to some degree to predict missing phenotypical characteristics of viruses and therefore can be used in different schemes of virus classification. But the possibility to construct binomial classification only can be used for viruses themselves [20]. Remind that using the table form for representing virus groups instead of hierarchical trees has been discussed in the studies of Lubischew [17]. Note that natural system defined in [17] is such that most of the properties of elements should be defined by their position in it. What is also important, the table form can potentially reveal empty cells not occupied by known virus genomes. This will give the possibility to predict the existence of new viruses having new characteristics of periodicity, monotony and symmetry in replicator patterns. Of course, the VGT proposed in this paper is only a first approach to create table form for virus genomes and is far from the ideal of natural system for viruses. We hope, nevertheless, that it can be a useful tool in current metagenomics studies.

# References

[1] Simmonds P et al (2017) Virus taxonomy in the age of metagenomics. Nature Reviews Microbiology. 15: 161–168
[2] van Regenmortel MHV, Mahy BWJ (2004) Emerging Issues in Virus Taxonomy. Emerg. Infect Dis 10(1): 8–13
[3] Simmonds P, Aiewsakun P (2018) Virus classification – where do you draw the line? Archives of Virology 163:2037–2046
[4] King AMQ et al (2018) Changes to taxonomy and the International Code of Virus Classification and Nomenclature ratified by the International Committee on Taxonomy of Viruses. Arch Virol. 163 (9): 2601–2631
[5] Pringle CR (1991). The 20th Meeting of the Executive Committee of ICTV. Virus species, higher taxa, a universal database and other matters. Arch Virol. 119: 303–4
[6] Hopfield JJ (1982) Neural networks and physical systems with emergent collective computational abilities. PNAS 79: 2554-2558
[7] Ezhov AA, Vvedensky VL (1996) Object Generation with Neural Networks (When Spurious Memories are Useful). Neural Networks. 9(9): 1491-1495
[8] Adams MJ, Lefkowitz EJ, King AMQ, Carstens EB (2013) Recently agreed changes to the international code of virus classification and nomenclature. Arch Virol. 158: 2633-2639
[9] Wakler PJ et al (2021) Changes to virus taxonomy and to the International Code of Virus Classifcation and Nomenclature ratifed by the International Committee on Taxonomy of Viruses (2021). Arch. Virol. 166:2633–2648
[10] Van Regenmortel MH, Ackermann HW, Calisher CH, Dietzgen RG, Horzinek MC, Keil GM, Mahy BW, Martelli GP, Murphy FA, Pringle C, Rima BK, Skern T, Vetten HJ, Weaver SC (2013) Virus species polemics: 14 senior virologists oppose a proposed change to the ICTV definition of virus species. Arch. Virol. 158(5): 1115-1119
[11] Calisher CH, Horzinek MC, Mayo MA, Ackermann HW, Maniloff J (1995) Sequence analyses and a unifying system of virus taxonomy: consensus via consent. Arch Virol 140:2093–2099
[12] Ball LA (2005) The universal taxonomy of viruses in theory and practice. In: Fauquet CM, Mayo MA, Maniloff J, Desselberger U, Ball LA (eds) Eighth ICTV Report. Elsevier, Amsterdam, pp 11–1
[13] GJ Morgan GJ (2016) What is virus species? Radical pluralism in viral taxonomy. Studies in History and Philosophy of Science, Part C: Studies in History and Philosophy of Biological and Biomedical Sciences. 59: 64-70
[14] Edgar RC (2004) Muscle: multiple sequence alignment with high accuracy and high throughput. Nucleic Acids Res. 32: 1792-1797
[15] T. Chookajorn (2020) Evolving COVID-19 conundrum and its impact Proc. Natl. Acad. Sci. U.S.A., 117: 12520-12521
[16] Lily He , Siyang Sun , Qianyue Zhang , Xiaona Bao , Peter K. Li (2021) Alignment-free sequence comparison for virus genomes based on location correlation coefficient  Infection, Genetics and Evolution 96: 105106.Available online 6 October 2021 1567-1348.
[17] Lubischew AA (1963). On some contradictions in general taxonomy and evolution. Evolution. 17(4): 414–430
[18] Bo Xia and Itai Yanai (2019). A periodic table of cell types. The Company of Biologists Ltd | Development 146
[19] Ezhov AA (2020) Can artificial neural replicators be useful for studying RNA replicators? Archives of Virology 165(11): 2513-2529
[20] van Regenmortel MHV (2007) Virus species and virus identification: Past and current controversies. Infection, Genetics and Evolution 7(1): 133-144
[21] Wakler PJ et al (2021). Changes to virus taxonomy and to the International Code of Virus Classification and Nomenclature ratifed by the International Committee on Taxonomy of Viruses. Arch. Virol. 166:2633–2648
[22] Ezhov AA, Vvedensky VL, Khromov AG, Knizhnikova LA (1991) Self-reproducible neural networks with synchronously changing neuronal threshold. In: Holden AV, Kryukov VI (eds) Neurocomputers and attention II: connectionism and neurocomputers, Manchester University Press, pp 523- 534
[23] Ezhov AA, Khromov AG, Knizhnikova LA, Vvedensky VL (1991) Self-reproducible networks: classification, antagonistic rules and generalization. Neural Networks World 1: 52–57
[24] Flores R, Ruiz-Ruiz S, Serra P (2012) Viroids and hepatitis delta virus. Semin Liver Dis. 32(3): 201-210
[25] Van Regenmortel MHV, Fauquet CM, Bishop DHL, Calisher CH, Carsten EB, Estes MK, Lemon, SM, Maniloff J, Mayo MA, McGeoch DJ, Pringle CR, Wickner RB (2002) Virus Taxonomy. Seventh Report of the International Committee for the Taxonomy of Viruses" Academic Press, New-York, San Diego
[26] Gheit T (2019) Mucosal and Cutaneous Human Papillomavirus Infections and Cancer Biology. Front. Oncol. 9:355
[27] Bzhalava D, Eklund C, Dillner J (2015) International standardization and classification of human papillomavirus types. Virology 476: 341-344
[28] Bruni L, Albero G, Serrano B, Mena M, Collado JJ, Gómez D, Muñoz J, Bosch FX, de Sanjosé S. (2021) ICO/IARC Information Centre on HPV and Cancer (HPV Information Centre). Human Papillomavirus and Related Diseases in the World. Summary Report.
[29] Van Doorslaer et al (2018) ICTV Virus Taxonomy Profile: Papillomaviridae. Journal of General Virology 99: 989–990
[30] La Rosa G (2016). Papillomavirus. In: Rose JB and Jiménez-Cisneros B, (eds) Water and Sanitation for the 21st Century: Health and Microbiological Aspects of Excreta and Wastewater Management (Global Water Pathogen Project). (Meschke JS, and Girones R (eds), Part 3: Specific Excreted Pathogens: Environmental and Epidemiology Aspects - Section 1: Viruses), Michigan State University, E. Lansing, MI, UNESCO
[31] Flores-Miramontes MG, Olszewski D, Artaza-Irigaray C, Willemsen A, Bravo IG, Vallejo-Ruiz V, Leal-Herrera YA, Piña-Sánchez P, Molina-Pineda A, Cantón-Romero JC, Martínez-Silva MG, Jave-Suárez LF and Aguilar-Lemarroy A (2020) Detection of Alpha, Beta, Gamma, and Unclassified Human Papillomaviruses in Cervical Cancer Samples From Mexican Women. Front. Cell. Infect. Microbiol. 10:234.
[32] Tampa M, Mitran CI, Mitran MI, Nicolae I, Dumitru A, Matei C, Manolescu L, Popa GL, Caruntu C, Georgescu SR (2020). The Role of Beta HPV Types and HPV-Associated Inflammatory Processes in Cutaneous Squamous Cell Carcinoma. Hindawi Journal of Immunology Research 2020 1: 1-10
[33] Pastrana DV, Peretti A, Welch NL, Borgogna C, Olivero C, Badolato R, et al. Metagenomic discovery of 83 new human papillomavirus types in patients with immunodeficiency. mSphere. 2018; 3:e00645.

[34] Doorbar J, Egawa N, Griffin H, Kranjec C and Murakami I (2016) Human papillomavirus molecular biology and disease association. Reviews in Medical Virology 25: 2–23
[35] Hirt L, Hirsch-Behnam A, de Villiers EM (1991) Nucleotide sequence of human papillomavirus (HPV) type 41: an unusual HPV type without a typical E2 binding site consensus sequence. Virus Research 18: 179-189
[36] Van Doorslaer K (2013) Evolution of the Papillomaviridae. Virology 445: 11-20
[37] Shah SD, Doorbar J, Goldstein RA (2010) Analysis of Host–Parasite Incongruence in Papillomavirus Evolution Using Importance Sampling. Molecular Biology and Evolution 27(6): 1301-1314
[38] Calvignac-Spencer S et al (2016) A taxonomy update for the family Polyomaviridae. Arch Virol. 161(6):1739-50
[39] Bhat AI, Hohn T, Selvarajan R (2016). Badnaviruses: The current global scenario. Viruses 8:177-205
[40] Remans T et al (2007) Banana Streak Virus: a Highly Diverse Plant Pararetrovirus Plant Viruses 1(1), 33-38
[41] Parisi G (2006) Spin glasses and fragile glasses: statics, dynamics, and complexity. PNAS 103(21): 7948-7955

**Acknowledgments** I am very grateful to Professor Marc H. V. van Regenmortel for his interest to our attempts to apply neural networks to genome analysis and also to Professors Eörs Szathmáry and Nikolay A. Makarenko for their interest to our model of neural replicators.

**Materials**

Below are the data used in the study of human papillomaviruses (Section 2): they contain the species name, type, NCBI and GenBank accession number, and the length of the virus dsDNA.

Genus: *Alphapapillomavirus*

| Species | Type | Accession | Length | Species | Type | Accession | Length |
|---|---|---|---|---|---|---|---|
| α-1 | HPV32 | NC_001586.1 | 7961 bp | α-7 | HPV18 | LC636309 | 7857 bp |
| | HPV42 | LR862086 | 7920 bp | | HPV39 | LR862071 | 7833 bp |
| α-2 | HPV3 | X74462.1 | 7820 bp | | HPV45 | EF202167.1 | 7849 bp |
| | HPV10 | X74465 | 7919 bp | | HPV59 | LR862080.1 | 7898 bp |
| | HPV28 | U31783.1 | 7959 bp | | HPV68 | GQ472851.1 | 7830 bp |
| | HPV29 | U31784.1 | 7916 bp | | HPV70 | U21941.1 | 7905 bp |
| | HPV77 | Y15175 | 7887 bp | | HPV97 | EF436229.1 | 7843 bp |
| | HPV78 | AB793779 | 7805 bp | α-8 | HPV7 | MK463913 | 8037 bp |
| | HPV94 | GU117628 | 7872 bp | | HPV40 | X74478 | 7909 bp |
| | HPV117 | GQ246950.1 | 7895 bp | | HPV43 | LR861953 | 8007 bp |
| | HPV125 | FN547152.2 | 7809 bp | | HPV91 | AF419318.1 | 7966 bp |
| | HPV160 | AB745694 | 7779 bp | α-9 | HPV16 | NC_001526.4 | 7906 bp |
| α-3 | HPV61 | U31793.1 | 7989 bp | | HPV31 | LR862053 | 7878 bp |
| | HPV62 | AY395706.1 | 8092 bp | | HPV33 | M12732.1 | 7909 bp |
| | HPV72 | X94164.1 | 7988 bp | | HPV35 | M74117.1 | 7851 bp |
| | HPV81 | AJ620209.1 | 8070 bp | | HPV52 | LC373207.1 | 7906 bp |
| | HPV83 | AF151983 | 8104 bp | | HPV58 | LC376008 | 7824 bp |
| | HPV84 | AF293960 | 7948 bp | | HPV67 | D21208 | 7801 bp |
| | HPV86 | AF349909 | 7983 bp | α-10 | HPV6 | AF092932 | 8012 bp |
| | HPV87 | KU298941.1 | 7992 bp | | HPV11 | HE574705 | 7933 bp |
| | HPV89 | KU298945.1 | 8074 bp | | HPV13 | X62843 | 7880 bp |
| | HPV102 | DQ090083.1 | 8078bp | | HPV44 | LR862067 | 7836 bp |
| | HPV114 | GQ244463.1 | 8069 bp | | HPV74 | LR862050 | 7902 bp |
| α-4 | HPV2 | MN605988.1 | 7859 bp | α-11 | HPV34 | KF436817 | 7788 bp |
| | HPV27 | AB211993.1 | 7831 bp | | HPV73 | LR862011 | 7716 bp |
| | HPV57 | MK463925 | 7848 bp | α-13 | HPV54 | HPU37488 | 7759 bp |
| α-5 | HPV26 | NC_001583.1 | 7855 bp | α-14 | HPV71 | NC_039089 | 8017 bp |
| | HPV51 | KF436884 | 7815 bp | | HPV90 | NC_004104 | 8033 bp |
| | HPV69 | KF436864.1 | 7705 bp | | HPV196 | DQ080082 | 8035 bp |
| | HPV82 | AB027021.1 | 7821 bp | | | | |
| α-6 | HPV30 | LR862000 | 7843 bp | | | | |
| | HPV53 | NC_001593.1 | 7856 bp | | | | |
| | HPV56 | LR862083 | 7866 bp | | | | |
| | HPV66 | LC511686.1 | 7818 bp | | | | |

Genus: *Betapapillomavirus*

| Species | Type | Accession | Length | Species | Type | Accession | Length |
|---|---|---|---|---|---|---|---|
| β-1 | HPV5 | JN211194 | 7746 bp | | HPV23 | U31781.1 | 7324 bp |
| | HPV8 | M12737.1 | 7654 bp | | HPV37 | U31786.1 | 7421 bp |
| | HPV12 | X74466.1 | 7673 bp | | HPV38 | JN211196 | 7397 bp |
| | HPV14 | X74467.1 | 7439 bp | | HPV80 | Y15176.1 | 7427 bp |
| | HPV19 | X74470.1 | 7685 bp | | HPV100 | FM955839.1 | 7380 bp |
| | HPV20 | U31778.1 | 7757 bp | | HPV104 | FV955840 | 7386 bp |
| | HPV21 | U31779.1 | 7779 bp | | HPV107 | EF42222.1 | 7562 bp |
| | HPV24 | U31782.1 | 7452 bp | | HPV110 | EU410348.1 | 7423 bp |
| | HPV25 | X74471.1 | 7713 bp | | HPV111 | EU410349.1 | 7384 bp |
| | HPV36 | U31785.1 | 7722 bp | | HPV113 | FM955842.1 | 7412 bp |
| | HPV47 | M32305.1 | 7726 bp | | HPV120 | FN598907 | 7304 bp |
| | HPV93 | AY382778 | 7450 bp | | HPV122 | GQ845444.1 | 7397 bp |

|  | HPV98 | FM955837.2 | 7466 bp |  | HPV145 | HM999997 | 7375 bp |
|---|---|---|---|---|---|---|---|
|  | HPV99 | FM955838 | 7698 bp |  | HPV151 | FN77756 | 7386 bp |
|  | HPV105 | FM955841.1 | 7667 bp |  | HPV159 | HE963025 | 7443 bp |
|  | HPV118 | GQ246951.1 | 7597 bp |  | HPV174 | HF930491.1 | 7359 bp |
|  | HPV124 | GQ845446.1 | 7489 bp | β-3 | HPV49 | NC_001591.1 | 7560 bp |
|  | HPV143 | HM999995 | 7715bp |  | HPV75 | Y15173.1 | 7537 bp |
|  | HPV152 | JF304768 | 7480 bp |  | HPV76 | Y15174 | 7549 bp |
| β-2 | HPV9 | NC_001596.1 | 7434 bp |  | HPV115 | FJ947080.1 | 7476 bp |
|  | HPV15 | X74468.1 | 7412 bp | β-4 | HPV92 | NC_004500.1 | 7461 bp |
|  | HPV17 | JN211195 | 7426 bp | β-5 | HPV96 | NC_005134.2 | 7438 bp |
|  | HPV22 | U31780.1 | 7368 bp |  | HPV150 | FN677755.1 | 7336 bp |

Genus: *Gammapapillomavirus*

| *Species* | Type | Accession | Length | *Species* | Type | Accession | Length |
|---|---|---|---|---|---|---|---|
| γ-1 | HPV4 | NC_001457.1 | 7353 bp | γ-11 | HPV126 | NC_016157.1 | 7326 bp |
|  | HPV65 | X70829.1 | 7308 bp |  | HPV136 | NC_017994.1 | 7319 bp |
|  | HPV95 | AJ620210.1 | 7337 bp |  | HPV140 | NC_017996.1 | 7341 bp |
|  | HPV158 | KT698168.1 | 7192 bp |  | HPV141 | HM999993 | 7384 bp |
|  | HPV173 | KF006400.1 | 7297 bp |  | HPV154 | NC_021483.1 | 7286 bp |
|  | HPV205 | KT698167.1 | 7298 bp |  | HPV169 | JX413105.1 | 7252 bp |
| γ-2 | HPV48 | NC_001690.1 | 7100 bp |  | HPV171 | KF006398.1 | 7261 bp |
|  | HPV200 | KP692114.1 | 7137 bp |  | HPV202 | KP692116.1 | 7344 bp |
| γ-3 | HPV50 | NC_001691.1 | 7184 bp | γ-12 | HPV127 | NC_014469.1 | 7181 bp |
| γ-4 | HPV60 | NC_001693.1 | 7313 bp |  | HPV132 | NC_014955.1 | 7125 bp |
| γ-5 | HPV88 | NC_010329.1 | 7326 bp |  | HPV148 | GU129016.1 | 7164 bp |
| γ-6 | HPV101 | LR861930 | 7259 bp |  | HPV157 | KT698166.1 | 7154 bp |
|  | HPV103 | NC_008188.1 | 7263 bp |  | HPV165 | JX444072.1 | 7129 bp |
|  | HP108 | NC_012213.1 | 7149 bp |  | HPV199 | KJ913662.1 | 7184 bp |
| γ-7 | HPV109 | NC_012485.1 | 7346 bp | γ-13 | HPV128 | NC_014952.1 | 7259 bp |
|  | HPV123 | GQ845445.1 | 7329 bp |  | HPV153 | JN171845 | 7240 bp |
|  | HPV134 | NC_014956.1 | 7309 bp | γ-14 | HPV131 | NC_014954.1 | 7182 bp |
|  | HPV138 | HM999990.1 | 7353 bp | γ-15 | HPV135 | NC_017993.1 | 7293 bp |
|  | HPV139 | HM999991.1 | 7360 bp |  | HPV146 | HM999998 | 7265 bp |
|  | HPV149 | GU117629.1 | 7333 bp |  | HPV179 | NC_022095.1 | 7228 bp |
|  | HPV155 | JF906559.1 | 7352 bp | γ-16 | HPV137 | NC_017995.1 | 7236 bp |
|  | HPV170 | JX413110.1 | 7417 bp | γ-17 | HPV144 | NC_017997.1 | 7271 bp |
| γ-8 | HPV112 | NC_012486.1 | 7227 bp | γ-18 | HPV175 | NC_038524.1 | 7226 bp |
|  | HPV119 | GQ845441.1 | 7251 bp | γ-19 | HPV161 | NC_038522.1 | 7238 bp |
|  | HPV147 | HM999996.1 | 7224 bp |  | HPV162 | JX413108.1 | 7214 bp |
|  | HPV164 | JX413106.1 | 7233 bp |  | HPV166 | NC_019023.1 | 7212 bp |
|  | HPV168 | KC862317.1 | 7204 bp | γ-20 | HPV163 | NC_028125.1 | 7233 bp |
| γ-9 | HPV116 | NC_013035.1 | 7184 bp | γ-21 | HPV167 | NC_022892.1 | 7228 bp |
|  | HPV129 | NC_014953.1 | 7219 bp | γ-22 | HPV172 | NC_038523.1 | 7203 bp |
| γ-10 | HPV121 | NC_014185.1 | 7342 bp | γ-23 | HPV156 | NC_033781.1 | 7329 bp |
|  | HPV130 | GU117630.1 | 7388 bp | γ-24 | HPV178 | NC_023891.1 | 7314 bp |
|  | HPV133 | GU117633.1 | 7358 bp |  | HPV197 | KM085343 | 7278 bp |
|  | HPV142 | HM999994.1 | 7374 bp | γ-25 | HPV184 | NC_038914.1 | 7324 bp |
|  | HPV180 | KC108722.1 | 7356 bp | γ-27 | HPV207 | MK645900.1 | 7247 bp |

Genera: *Mu-, Nu- and purcupine Sigma- papillomaviruses*

| μ-1 | HPV1 | NC_001356.1 | 7815 bp |
|---|---|---|---|
| μ-2 | HPV63 | NC_001458.1 | 7348 bp |
| μ-3 | HPV204 | NC_038525.1 | 7227 bp |
| ν | HPV41 | NC_001354.1 | 7614 bp |
| σ | EdPV-1 | NC_006951.1 | 7428 bp |